\documentclass[twocolumn,iop,apj]{emulateapj} %%emulateapj
\usepackage{amsfonts,amsmath,amssymb,epsfig}
\usepackage{graphicx,dcolumn,bm}
\usepackage{mathptmx}                    % which font to use ?
\usepackage[utf8]{inputenc}
\usepackage{subfigure}

\def\mth{M_{\mathrm{threshold}}^{H}}
\def\mtq{M_{\mathrm{threshold}}^{Q}}
\def\mhyb{M_{\mathrm{TOV}}^{\mathrm{Coll-hyb}}}
\def\mthyb{M_{\mathrm{threshold}}^{\mathrm{Coll-hyb}}}
\def\mshyb{M_{\mathrm{supra}}^{\mathrm{Coll-hyb}}}
\def\msup{M_{\mathrm{supra}}}
\def\ms{M_{\odot}}
\def\mtov{M_\text{TOV}}

\def\tl{\tilde\Lambda}
\def\mtot{M_{\mathrm{tot}}}
\def\mrem{M_{\mathrm{remnant}}}
\def\mhmax{{M^H_\mathrm{max}}}

\newcommand{\codename}[1]{\texttt{#1}}

\begin{document}

\title{Merger of compact stars in the two-families scenario}

\author{
Roberto De Pietri $^{1,2}$,
Alessandro Drago$^{3,4}$, Alessandra Feo$^{5,2}$, \\Giuseppe Pagliara$^{3,4}$, 
Michele Pasquali$^{1,2}$, Silvia Traversi$^{3,4}$, Grzegorz Wiktorowicz$^{6,7}$}

\affiliation{
$^1$ Parma University, Parco Area delle Scienze 7/A, I-43124 Parma, Italy\\
$^2$ INFN gruppo collegato di Parma, Parco Area delle Scienze 7/A, I-43124 Parma, Italy\\
$^3$ Dipartimento di Fisica e Scienze della Terra dell'Universit\`a di Ferrara, Via Saragat 1, I-44100 Ferrara, Italy\\
$^4$ INFN Sez.~di Ferrara, Via Saragat 1, I-44100 Ferrara, Italy\\
$^5$ Department of Chemistry, Life Sciences and Environmental Sustainability,
Parma University, Parco Area delle Scienze, 157/A, I-43124 Parma (PR), Italy \\
$^6$ National Astronomical Observatories, Chinese Academy of Sciences, 
Beijing 100012, China\\ 
$^7$ China and School of Astronomy and Space Science, University of the Chinese Academy of Sciences, 
Beijing 100012, China}

\begin{abstract}
We analyse the phenomenological implications of the two-families
scenario on the merger of compact stars. That scenario is based on the
coexistence of both hadronic stars and strange quark stars. After
discussing the classification of the possible mergers, we turn to
detailed numerical simulations of the merger of two hadronic stars,
i.e., "first family" stars in which delta resonances and hyperons are
present, and we show results for the threshold mass of such binaries,
for the mass dynamically ejected and the mass of the disk surrounding
the post-merger object. We compare these results with those obtained
within the one-family scenario and we conclude that relevant
signatures of the two-families scenario can be suggested, in
particular: the possibility of a rapid collapse to a black hole for
masses even smaller than the ones associated to GW170817; during the
first milliseconds, oscillations of the postmerger remnant at
frequencies higher than the ones obtained in the one-family scenario;
a large value of the mass dynamically ejected and a small mass of the
disk, for binaries of low total mass.  Finally, based on a population
synthesis analysis, we present estimates of the number of mergers for:
two hadronic stars; hadronic star - strange quark star; two strange
quark stars. We show that for unequal mass systems and intermediate
values of the total mass, the merger of a hadronic star and a strange
quark star is very likely (GW170817 has a possible interpretation into
this category of mergers). On the other hand, mergers of two strange
quark stars are strongly suppressed.
\end{abstract}

\maketitle

%===============================================================================
\section{Introduction}
The first detection of gravitational waves (GW) from the merger of two
compact stars\footnote{Here we use the generic name compact stars to
  indicate either neutron stars (NSs), hadronic stars HSs or strange
  quark stars QSs.} in August 2017 \citep{TheLIGOScientific:2017qsa}
represents a breakthrough for the astrophysics of compact objects and
the physics of dense nuclear matter. The GW signal associated with the
process of inspiral of two stars encodes information on the average
tidal deformability $\tl$ of the binary which is strongly dependent on
the stiffness of the equation of state (EoS) of dense matter. Several
recent studies have exploited the limits on $\tl$, to obtain
constraints on the radii of compact stars
\citep{Fattoyev:2017jql,Most:2018hfd,Lim:2018bkq,Abbott:2018exr,Burgio:2018yix}.
The electromagnetic counterparts of GW170817 i.e., the short
gamma-ray-burst (sGRB) GRB170817A and the kilonova (KN) AT2017gfo,
allow to further constrain the EoS by indicating that the most
probable merger's remnant is a hypermassive compact star and by
requiring that the mass released and powering the KN is of the order
of $0.05M_{\odot}$
\citep{Bauswein:2017vtn,Margalit:2017dij,Annala:2017llu,Ruiz:2017due,Radice:2017lry,Rezzolla:2017aly}. The
main conclusion of those studies can be summarized by stating that the
EoS of dense matter cannot be very stiff: EoSs leading to radii larger
than about $13.5$ km are basically ruled out; moreover the maximum
mass of the non-rotating configuration should be less than
approximately $\sim 2.2\ms$ although still with a large error bar of
the order of $0.2\ms$. This is an interesting result: indications of
radii larger than approximately $13.5$km would indeed point to a stiff
EoS in which only nucleons are present while for smaller radii
non-nucleonic degrees of freedom could appear in compact stars.

The relevant degrees of freedom for the composition of a compact star
can be inferred from the value of the radius of a star having a mass
of about $1.5 \ms$. Firstly, if the radius is large, 13 km$\lesssim
R_{1.5}\lesssim$13.5 km the density at the center is low enough that
non-nucleonic degrees of freedom appear only in the most massive stars
\citep{Lonardoni:2014bwa}. Secondly, if 11.5 km$\lesssim
R_{1.5}\lesssim$13 km resonances (hyperons and/or deltas) will appear
even for stars having masses of the order of $\sim 1.5 \ms$
\citep{Maslov:2015msa}. Also deconfined quarks can appear at the
center of the star which becomes a so-called hybrid star
\citep{Nandi:2017rhy}. Finally, for $R_{1.5}\ll 11.5$km only
disconnected-solutions of the Tolman-Oppenheimer-Volkov equation exist
\citep{Alford:2013aca,Burgio:2018yix}. These solutions are
characterized by the presence of a mass-window inside which two
different configurations are possible: one made only of hadronic
matter and the other made partially or totally of deconfined quark
matter. These configurations correspond either to the twin-stars
scenario \citep{Alford:2013aca,Paschalidis:2017qmb,Most:2018hfd} or to
the two-families scenario
\citep{Drago:2013fsa,Drago:2015cea,Drago:2015dea,Wiktorowicz:2017swq}.

The main differences between these two scenarios concern the
mass-radius relation and the microphysics embedded in the EoS.  In the
twin-stars scenario the stars containing quarks are both the most
massive and the ones with the smallest radius, which reaches its
minimum at the maximum mass configuration. On the other hand, in the
two-families scenario, the stars having the smallest radius belong to
the hadronic branch (Hadronic Stars, HSs) and they have a mass of the
order of $(1.5-1.6) \ms$. Instead, the second family, which
corresponds to Strange Quark Stars (QSs), is characterized by not too
small radii (of the order or larger than about $12$km) and a maximum
mass $\mtov^Q$ that can exceed $2\ms$.  Concerning the EoS: the
twin-stars scenario assumes the existence of a mixed phase of hadrons
and quarks whereas in the two-families scenario one assumes the
validity of the Bodmer-Witten hypothesis
\citep{Bodmer:1971we,Witten:1984rs} which implies the existence of a
global minimum of the energy per baryon at a finite density and for a
finite value of strangeness.

The investigation of these scenarios as well as their consequences on
the interpretation of GW170817 has been discussed in previous papers
\citep{Paschalidis:2017qmb,Burgio:2018yix,Nandi:2017rhy,Gomes:2018eiv}.
In particular, in \citet{Burgio:2018yix} it has been shown that the
observational constraints on $\tl$ would imply that the radius of the
$1.4\ms$ configuration, $R_{1.4}$, is larger than about $12$ km within
the one family scenario, while when considering two families of
compact stars $R_{1.4}$ could be significantly smaller, down to about
$10$ km.

 In this paper, we extend a previous preliminary work on the
 predictions of HS-HS mergers within the two families scenario
 \citep{Drago:2017bnf}. After an overview of the phenomenology of
 mergers within the two-families scenario, we present numerical
 simulations of the process of merger of two HSs by using the {\tt
   Einstein toolkit}, an open source, modular code for numerical
 relativity \citep{Loffler:2011ay}. We consider two EoSs: a soft one
 which contains hyperons and deltas and leads to HSs belonging to the
 first family and, as a reference, a stiffer and purely nucleonic
 EoS. The main outcome of the simulations is a precise estimate of the
 threshold mass above which a prompt collapse occurs for the merger of
 two HSs. We also provide estimates of the mass ejected which is an
 important quantity for the phenomenology of KNe. Also, we discuss in
 which region of the star quark matter nucleation can take place.
 Finally, we estimate the various possible merger processes within the
 two families scenario by using the population synthesis code {\tt
   Startrack}
 \citep{Belczynski0206,Belczynski0801,Wiktorowicz:2017swq}.

\section{Equations of state of hadronic matter and quark matter}\label{sec:EOS}
The two families scenario is based on the idea that in the hadronic
EoS delta resonances and hyperons do appear at large densities. To
model the EoS we adopt the relativistic mean field model SFHo of
\citet{Steiner:2012rk} with the inclusion of delta resonances and
hyperons, see \citet{Drago:2014oja}. Here we use the EoS SFHo-HD
already investigated in \citet{Burgio:2018yix} and which corresponds
to a coupling of delta resonances with the sigma meson of $x_{\sigma\Delta}=1.15$ (see \citet{Drago:2014oja,Burgio:2018yix} for
details).

For quark matter we use the simple bag-like parametrizations of \citet{Alford:2004pf,Weissenborn:2011qu} where an effective bag constant $B_{\mathrm{eff}}$ has been introduced together with a parameter $a_4$ which encodes pQCD corrections.
By setting $B_{\mathrm{eff}}^{1/4}=137.5$ MeV, $a_4=0.75$ and the mass of the strange quark $m_s=100$ MeV we obtain 
$\mtov^Q \sim 2.1\ms$.

\subsection{Transition from hadronic matter to quark matter}
To model the formation of quark matter we adopt the scheme based on
quantum nucleation developed in many papers, see e.g.,
\citet{Iida:1998pi,Berezhiani:2002ks,Bombaci:2004mt}.  The basic idea
is that the process of formation of the first drop of quark matter
preserves the flavor composition and it is therefore impossible to
nucleate strange quark matter if hyperons or kaons are not already
present in the hadronic phase.  In Fig.1 we show the fraction $Y_i$ of
baryons as a function of the baryon density $n_b$.  Note that, as
$n_b$ increases, more and more hyperons are produced making the
conversion to strange quark matter more likely.  Let us define the
strangeness fraction $Y_S=(n_{\Lambda}+2(n_{\Xi^0}+n_{\Xi^-}))/n_b$.
A possible way to define a threshold for the conversion of hadronic
matter into strange quark matter is to request that $d_s$, the average
distance of strange quarks inside the confined phase, is of the order
or smaller than the average distance of nucleons in nuclear matter
$d_n=n_0^{-1/3}$, where $n_0$ is the nuclear matter saturation
density, $n_0=0.16$fm$^{-3}$. The reason is that strange quarks should
be close enough to be able to mutually interact in order to help the
nucleation of a first drop of deconfined quark matter. This way of
defining a critical density is simple and intuitively it corresponds
to a necessary condition but it is less quantitative than the one
based on discussing the minimal size of the stable drop of quark
matter used in many papers, including the ones cited above.  One
should also notice that at the high temperature reached during the
merger, thermal nucleation could be possible and it could make the
whole process of quark deconfinement significantly faster (e.g. in
\citet{DiToro:2006bkw} the temperature above which thermal nucleation
becomes dominant has been estimated to be around 60 MeV).

\begin{figure}[!ht]
	\begin{centering}
		\epsfig{file=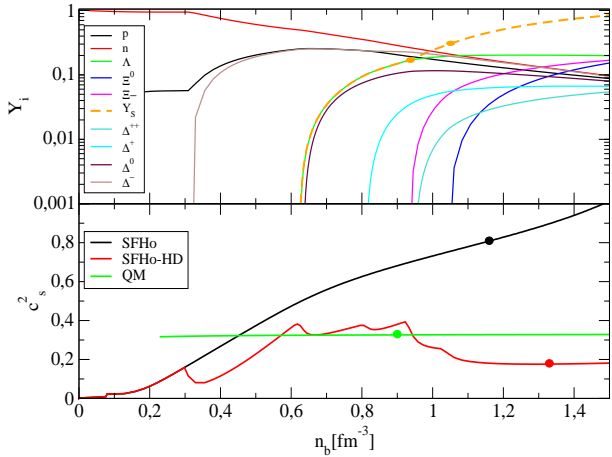,height=8cm,width=6cm,angle=-90}
		\caption{Upper panel: particle fractions and strangeness fraction $Y_S$ as functions of the baryon density for beta-stable matter at zero temperature. The orange points indicate the range of densities and strangeness fraction for which we expect nucleation of strange quark matter to take place: $0.2\lesssim Y_S \lesssim 0.3$ .\label{ys} Lower panel: density dependence of the speed of sound for the SFHo EoS, for its generalization which includes delta baryons and hyperons SFHo-HD and for the quark matter EoS (labelled QM). The filled dots indicate the central densities and the corresponding values of the speed of sound at the center of the maximum mass configuration. }
	\end{centering}
\end{figure}

In Fig.\ref{ys}, upper panel, we show the fraction of the strange and
not-strange baryons, together with the fraction of strangeness, at
zero temperature. The previous condition on the minimal density of
strangeness is satisfied above a baryon density of the order of
(0.9--1) $\mathrm{fm}^{-3}$ for which $Y_S\sim 0.2-0.3$ (see the two
points on the dashed line in the figure) and which corresponds to
about (6--7) $n_0$. Note that it is a density significantly larger
than the threshold density of formation of hyperons, which slightly
exceeds 0.6 $\mathrm{fm}^{-3}$ and from this viewpoint our conditions
agrees with the analysis of e.g. \citet{Bombaci:2004mt,Bombaci:2009jt}
indicating that the first droplet of quark matter is nucleated at
densities larger than the threshold density of hyperon formation. It
is important to notice that the condition for quark nucleation at
$T=0$ is reached at densities smaller than those corresponding to the
maximum mass mechanically stable, $\mtov^H$. Following
\citet{Bombaci:2004mt} we therefore introduce a separate notation
$\mhmax$ for the maximum mass of a hadronic star, stable respect to
quark nucleation.

A specific feature of the two families scenario and of the
quark-deconfinement mechanism is the possibility to form also (hot)
hybrid stars (HybSs) as a result of the partial turbulent conversion
of hadronic matter into quark matter. These stellar objects live for a
time of the order of $10$s, which is necessary for hadronic matter to
convert completely into quark matter within the diffusive regime
\citep{Drago:2015fpa} and they are characterised by a two-phases
structure with a hot quark matter core and a hadronic matter layer
(initially cooler) which are in mechanical equilibrium but out of
thermal and chemical equilibrium. It is possible to construct a model
for the EoS describing these transient HybSs by using the Coll's
condition \citep{Coll:1976} to define the density separating the low
density hadronic region from the high density region already made of
quarks, see \citet{Drago:2015fpa} for details. The reason to discuss
here these shortly living configurations is that the postmerger
remnant, for the cases of HS-HS and HS-QS mergers, can indeed form
such HyBs in which the conversion is still proceeding (we label it
Coll-hyb).  Notice that since QSs are generated in this scenario by
the conversion of HSs, they have a minimum mass ($M^Q_{\mathrm{min}}$)
which is fixed by the maximum mass of HSs: the conservation of baryon
number during the conversion implies that the total baryon number of
the newly born QS is equal to the total baryon number of the
progenitor HS. Since the QS is more bound than its progenitor HS, its
gravitational mass is of the order of $0.1\ms$ smaller than the
gravitational mass of the progenitor and its minimum value is of $\sim
(1.35-1.45)\ms$ depending on the value of the maximum mass of HSs, see
\citet{Wiktorowicz:2017swq}.

Another distinctive feature of the two-families scenario concerns the
value of the (square) speed of sound $c^2_s$ in dense matter: in the
hadronic EoS its value remains significantly below $\sim 0.4$ due to
the several softening channels associated with the appearance of
hyperons and deltas (see red line in the lower panel of Fig.1). For
the quark matter EoS, within the model here adopted, at increasing
density it approaches from below the asymptotic limit of the
non-interacting massless gas $1/3$ (the conformal limit), see green
line in the lower panel of Fig.1.  Nevertheless, it is possible to
fulfill the two solar mass limit by assuming that such massive stars
are QSs (see discussion in Sec.3.1).\footnote{Notice that the
  possibility to have massive QSs even for small values of the speed
  of sound was already recognised in \citet{Haensel:1986qb}, where it
  was noticed that the adiabatic index diverges at the surface of a
  QS, stabilizing that compact object.}  On the other hand, within the
one family scenario, the two solar mass limit requires values of the
speed of sound significantly larger than the asymptotic limit, as it
has been observed in \citet{Bedaque:2014sqa}.  For instance, by using
the SFHo EoS, one can obtain a maximum mass of $2.06\ms$
\citep{Steiner:2012rk} with a corresponding value of the central
density of $\sim 1.2$fm$^{-3}$ and $c^2_s \sim 0.8$, see the black
line in the lower panel of Fig.1 Similarly, within the twin star
scenario, the quark matter EoS is assumed to have $c^2_s \geq
0.8$\citep{Paschalidis:2017qmb}.  Since at asymptotically large
densities one should anyway recover the limit of $c^2_s=1/3$ one would
need to clarify which are the physical mechanisms responsible of an
initial increase of the speed of sound to values larger than about 0.8
and then to its decrease to the conformal limit.
 
\section{Static and rotating configurations}

The fate of the merger of two compact stars, in particular the
properties of the post-merger remnant and its emissions in GWs and
EMWs, is determined in general by two quantities: the total mass above
which a prompt collapse to a black hole (BH) takes place
$M_{\mathrm{threshold}}$ and the maximum mass of the supramassive
configuration $M_{\mathrm{supra}}$. Both quantities strongly depend on
the EoS.  The fact that the outcome of a merger is (mainly) determined
by those two quantities is true both in the one-family and in the
two-families scenario, but in the latter the situation is more
complicated because those quantities need to be determined for both
families (and also for the shortly-living Coll-hyb configuration).

Let us first present the results for the structure 
of static and rotating stars when adopting the EoSs previously introduced.
Several works have been
dedicated to the study of the dependence of $M_{\mathrm{threshold}}$
on the stiffness of the EoS: a very interesting result, presented in
\citet{Bauswein:2013jpa,Bauswein:2015vxa,Bauswein:2017aur} and based on
explicit numerical simulations, is that that ratio
$k=M_{\mathrm{threshold}}/\mtov$ scales linearly with the ratio
$\mtov/R_{\mathrm{TOV}}$, i.e. with the compactness of the maximum mass configuration. Depending on the EoS, $k$ varies between
$1.3$ and $1.6$. In Sec.5 we will discuss how to
estimate this parameter from direct numerical simulations of HS-HS mergers. Concerning
$M_{\mathrm{supra}}$, many studies have found that
$M_{\mathrm{supra}}\sim 1.2 \mtov$ \citep{Lasota:1995eu,Breu:2016ufb}
for the case of stars with a crust (i.e. HSs and HyBs) whereas for
QSs $M_{\mathrm{supra}}\sim 1.4 \mtov$
\citep{Gourgoulhon:1999vx,Stergioulas:2003yp}.

\begin{figure}[!ht]
	\begin{centering}
		\epsfig{file=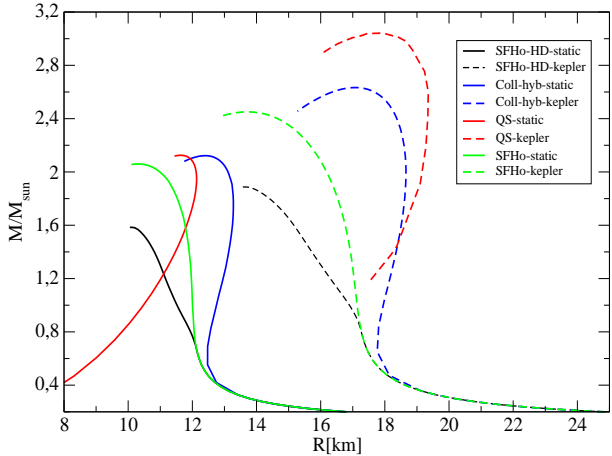,height=8cm,width=6cm,angle=-90}
		\caption{Static and keplerian mass-radius relations for HSs, QSs and Coll-HybSs. We show also the results for NSs described by the SFHo EoS (the results have been obtained by using RNS \citep{Stergioulas:1994ea,Stergioulas:2003yp}. \label{mrplot}}
	\end{centering}
\end{figure}

We have computed both static and keplerian
configurations by using the RNS code of \citet{Stergioulas:1994ea}.
Results are presented in Fig.\ref{mrplot}: for both HSs, QSs and NSs we confirm
the standard results previously described concerning the relation between $\mtov$ and $M_{\mathrm{supra}}$.
We have also computed $\mshyb$
for our Coll-hyb configurations which turns out to be of the order of $2.6M_{\odot}$, respecting the general relation between $\mtov$ and $M_{\mathrm{supra}}$ of compact stars with a crust. As discussed above, while Coll-hyb configurations are chemically and thermally out of equilibrium,
they represent a necessary intermediate stage of the evolution
of the post-merger remnant in the two-families scenario.
Their stability is therefore important both for 
HS-HS mergers and for HS-QS mergers. We have not computed the maximum mass of a differentially rotating Coll-hyb configuration $\mthyb$ through direct numerical simulations but we can safely assume that $\mthyb\geq \mshyb$.

For the sake of the following discussions let us summarize the possible values of the masses discussed above.
First, let us introduce the quantity $\mtot$  which is the sum of the gravitational masses of the two objects undergoing the merger: 
\begin{equation}
\mtot=M_1 + M_2.
\end{equation}
When discussing the critical masses one has to be careful to distinguish
between $\mtot$ and the gravitational mass that remains after mass ejection
(and GW emission) from the system, $\mrem$. The relation between these two
masses is of the form:
\begin{equation}
\mrem=(1-\alpha) \mtot \, ,
\end{equation}
where $\alpha$ is of the order of a few percent.
Plausible values of the masses discussed above are:
\begin{itemize}
    \item HSs:

    $\mtov^H \sim \mhmax \sim 1.6\ms$ \\ $M^H_{\mathrm{supra}} \sim 1.9\ms$ \\ $\mth \sim 2.5 \ms$
    
    \item QSs:\\ $\mtov^Q \sim 2.1\ms$\\ $M^Q_{\mathrm{supra}} \sim 3\ms$\\ $\mtq \sim M^Q_{\mathrm{supra}} \sim 3\ms$  \footnote{In \citet{Bauswein:2008gx} two quark EoSs have been investigated with $\mtov \sim 1.64 \ms$ and $1.88\ms$, respectively. From Fig.2 of that paper one can extract the ratios between $\mtq$ and $\mtov$ which turns out to be close to the ratio between $\msup$ and $\mtov$.}

    \item Coll-HyBs: \\
    $\mtov^{\mathrm{Coll-hyb}} \sim \mtov^Q \sim 2.1 \ms$\\
    $\mshyb \sim 2.6 \ms$ \\
    $M^{\mathrm{Coll-hyb}}_{\mathrm{remnant}}=(1-\alpha)\mthyb \sim  1.5\, \mtov^{\mathrm{Coll-hyb}} \sim 3.1\ms$
\end{itemize}

Concerning HSs, within the two-families scenario there is an uncertainty on $\mtov^H$ (whose value determines $M^H_{\mathrm{supra}}$ and $\mth$) and on $\mhmax$ due to the difficulty in estimating 
the critical value of the central density at which the formation 
of quark matter is triggered. A possible range of values for $\mhmax$ is 
$\mhmax=(1.5-1.6)\ms$ and $\mtov^H$ takes only slightly higher values. 

The value of $\mtov^Q$ has a larger uncertainty stemming from the 
unknown value of the maximum mass of compact stars which, 
as mentioned in the introduction, is estimated to be $(2.2\pm0.2)\ms$.
Interestingly, this same range is the one allowed by 
the microphysics of nucleation of quark matter in hadronic matter. For significantly larger values of $\mtov^Q$
the process of nucleation of quark matter would no more be 
energetically convenient \citep{proceeding}.

Finally, the uncertainty on $\mtov^Q$ translates into 
a similar uncertainty on the values of $\mtov^{\mathrm{Coll-hyb}}$.
Once a value for $\mtov^{\mathrm{Coll-hyb}}$ is chosen, the value of $\mshyb$ can been explicitly computed, while 
the value of $\mthyb$ has only been estimated by using the results of \citet{Weih:2017mcw}
indicating that the maximum mass of differentially rotating stars
is a factor $\sim 1.5$ larger than $\mtov$.

\section{Classification of the mergers}

\begin{figure}[!ht]
	\begin{centering}
		\epsfig{file=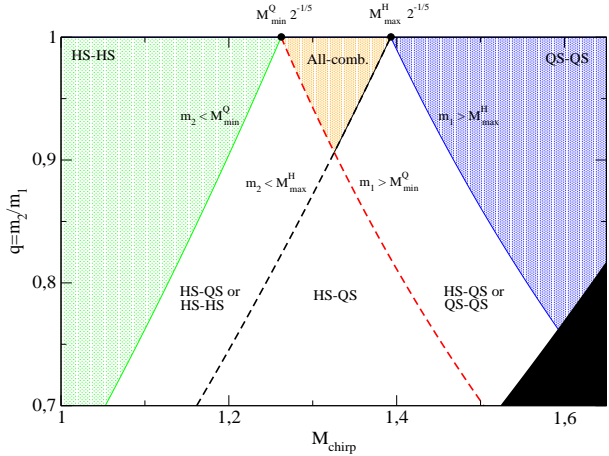,height=8cm,width=6cm,angle=-90}
		\caption{Different possible mergers within the two families scenario, depending on the chirp mass and on the mass asymmetry of the binary. Parameters: $\mhmax=1.6 \ms$, $\mtov^Q=2.1 \ms$ and the minimum masses of HSs ($M^H_{\mathrm{min}}$) and QSs ($M^Q_{\mathrm{min}}$) are set respectively to $1\ms$ and $1.45\ms$. The black area is excluded
		because there $m_1>M^Q_{\mathrm{TOV}}$.}
		\label{merger-cases}
	\end{centering}
\end{figure}

Within the two-families scenario there are three possible merger events: HS-HS, HS-QS and QS-QS. Notice that HS-HS is obviously the only possible type of merger in a one-family
scenario and that QS-QS has been discussed in the past in a few papers by
various authors without making any assumption on the number of families of compact stars \citep{Haensel:1991um,Bauswein:2008gx,Bauswein:2009im}. The HS-QS merger is instead possible only within the two-families
scenario.
These different possibilities are displayed 
in the diagram of Fig.\ref{merger-cases} in which the two axes correspond to the two parameters determining the main properties of the binary i.e., the   
chirp mass $M_{\mathrm{chirp}}$ and the mass asymmetry $q=m_2/m_1$ (for the sake of discussion we have made a specific choice of the maximum and minimum mass of HSs and QSs, see the caption).
It is clear how rich can be the phenomenology of mergers within the two-families scenario: for instance, within the small window labelled "All-comb." there could be events of HS-HS, HS-QS and QS-QS merger. On the other hand, there are regions of the diagram in which we predict only one specific type of merger to be possible (for instance only HS-HS mergers to the left of the green line). As a consequence, for a same value of $M_{\mathrm{chirp}}$, there could be a prompt collapse to a BH if the merger is of the HS-HS type while a hypermassive or a supramassive configuration could form if it is a HS-QS merger, see \citet{Drago:2017bnf}. Note that the classification presented in Fig.\ref{merger-cases} depends only on the maximum and minimum masses of the HSs and of the QSs since it does not suggest the outcome of the mergers (this question will be discussed in Sec.4.2), but it only indicates which types of mergers are possible within the two-families scheme. Finally, the probabilities of the various merger processes will be addressed in Sec.6 by using a population synthesis code.

\subsection{Quark deconfinement in the merger of compact stars}
To better understand why also the properties of the hybrid configurations Hyb-Coll are important to establish the fate of a merger 
in the two-families scenario,
one has to recall in which way the process
of quark deconfinement starts and proceeds inside a compact star. Quark deconfinement will evolve differently in the three types of merger we are discussing. 
\begin{itemize}
\item {\it HS-HS merger.}
In this case quark matter is not present in the system before the merger. As already discussed in the previous Section, the trigger
for quark matter nucleation is the presence of hyperons,
since they provide a finite density of strange quarks.
Once the density/temperature at the center of a
compact object exceed critical values at which the density of hyperons is large enough, the process of quark deconfinement starts. This implies
that quark deconfinement begins immediately after the merger
of two HSs, because few milliseconds after the merger the density and the temperature
increase and hyperons start forming at the center of the star \citep{Sekiguchi:2011mc}.
In Sec.5.6 we will investigate the conditions for quark nucleation by studying the results of the merger simulations. 
The process of quark deconfinement is initially very rapid,
since it is accelerated by hydrodynamical instabilities and in a few milliseconds it converts the bulk of the star into quarks. After this initial
phase the instabilities are suppressed and the process becomes much slower:
the conversion of the most external layer into quarks can take about ten
seconds.\footnote{Notice that the rapid burning stops at a energy density
which satisfies Coll's condition \citep{Drago:2015fpa}). Since the temperatures reached
by the system after the merger and therefore the energy density distribution
depend on the pre-merger configuration the value of $\mhyb$ is not "universal" but depends on the pre-merger history. The value we have provided is therefore just an approximation, the real value could be established through a simulation of the merger in which the process of quark deconfinement is described as
e.g. in the numerical simulation of \citet{Pagliara:2013tza}, but this is at the moment a very challenging numerical problem.} As done in Sec.3.1, it is therefore very important to estimate the maximum mass of the 
stable hybrid hadron-quark configuration that forms a few milliseconds after
the merger, $\mhyb$ and of its rotating counterparts $\mshyb$ and $\mthyb$, because
their values determine the fate of the newly formed object after the central
region has deconfined. 

\item {\it HS-QS merger.}
In this case quark matter is already present in the system. Notice that at the moment we are not able to simulate the merger HS-QS because it involves
two different EoSs. We can though compare this process with the HS-HS one,
by noticing that in the HS-QS case a large fraction of the star is already in the quark matter phase. The process of rapid combustion of hadrons into quarks can still take place if the quark matter phase does not occupy already the
whole region which satisfies Coll's condition. In this case, it is also possible that the process
of quark deconfinement is significantly faster, because the highly turbulent initial post-merger phase can mix quark matter and hadronic matter so that the mixing area is much larger than in the laminar case and the process is thus greatly accelerated. Also in this case $\mhyb$, $\mshyb$ and $\mthyb$ play an important role in the determination of the fate of the merger.

\item {\it QS-QS merger.}
This merger has been discussed in a few papers in the past (see for instance \citet{Bauswein:2009im}). The main reason
of the interest of this process is the possibility of a "clean" environment,
since it was assumed that little or no hadronic matter was emitted during and
after the merger \citep{Haensel:1991um}. Although this is not strictly true (quite a significant amount
of matter can be ejected dynamically at the moment of the merger) the system
is anyway more "clean" from baryon contamination since significantly less matter can be ejected by e.g. neutrino ablation. This is due to the much larger binding energy of strange quark matter respect to nuclear matter.
We will discuss later the possible phenomenological
signatures of this type of merger. It is clear that since in this case all hadronic matter is already made of deconfined quarks the only relevant
quantities are $\mtov^Q$, $\msup^Q$ and $\mtq$.
\end{itemize}

\subsection{Possible outcomes of a merger}

Let us discuss the possible outcomes of a merger with the associated phenomenology.
A first crucial problem we need to address is to
clarify in which way a sGRB can be produced in the two-families scenario.

\subsubsection{sGRB inner engine}
There are mainly two ways to generate a sGRB: one is based on the formation of a BH \citep{MacFadyen:1998vz} and the other on the formation of a proto-magnetar \citep{Metzger:2010pp}. These two mechanisms have something in common: they both need to wait for a time long enough so that the environment becomes less baryon loaded and a jet can be launched. The ultimate source of energy to power the sGRB is the accretion disk around the BH, in the first case, and the rotational energy in the case of the protomagnetar. If the sGRB is due to the formation of the BH the duration of the prompt emission is regulated by the duration of the disk around the BH: since the amount of material in the disk is significantly less in the case of mergers than in the case of collapsars one can explain why the duration of long GRBs and of sGRBs differ by roughly two orders of magnitude. It is more problematic to explain that difference within the protomagnetar mechanism, since the strength of the magnetic field and the rotational frequencies are comparable in the two cases \citep{Rowlinson:2013ue}. In \citet{Drago:2015qwa} it has been suggested that the duration of sGRB is linked to the time needed to deconfine the surface of the star close to the rotation axis: as long as nucleons are still present on that surface the baryonic load is too large to launch a jet, while when quarks occupy the whole surface the baryonic load rapidly drops to zero. The duration of the burst is therefore related to the time needed to deconfine completely the surface of the post-merger compact object.

In our scenario we assume that both mechanisms are possible: if a BH forms a few tens of ms after the merger (once differential rotation has been dissipated \footnote{The time needed for the dissipation of differential rotation is rather uncertain since it depends on the difficult estimates of the viscosity in the system. Recently it has been suggested that the hypermassive configuration could last for a time scale of a second \citep{Gill:2019bvq}.}), a sGRB can be launched by using the energy extracted from the accretion disk, while if the remnant is stable at least for a few seconds (as a supramassive star) it can generate a sGRB 
within the protomagnetar mechanism. In this second scenario, an extended emission could also be obtained with a duration 
which is connected to the time needed for the supramassive star to collapse \citep{Rowlinson:2013ue,Lu:2015rta}.

Let us discuss now the outcomes of the mergers
in the three possible cases within the two-families scenario. We will adopt for all the critical masses 
the values discussed in Sec.3.

\subsubsection{HS-HS merger.}
The HS-HS merger is the case examined in the numerical simulations of this paper. In the following sections we will determine the value of $\mth$ and we will investigate if the conditions for quark nucleation can be reached. Depending on the value of $\mtot$ the outcomes of the mergers are the following:

1) $\mtot > \mth$:  there is a direct collapse to BH. In this case we can expect to have a GW signal associated with the inspiral phase (with a very low value of $\tilde{\Lambda}$ since HSs with masses close to $\mhmax$ have very small radii), almost no GW signal in the post-merger, no sGRB and a very faint kilonova signal produced only by the mass ejected by tidal forces and by the shock wave (which is however small in this case, see discussion in Sec.5). \footnote{For highly asymmetric systems ($q \sim 0.8$), even in the cases of prompt collapse, a sGRB could be released due to the formation of a massive torus around the BH which could have a mass of a few $0.1\ms$, see \citet{Rezzolla:2010fd,Giacomazzo:2012zt}. Since 
the torus mass scales linearly with $\mtov$ (which is $\sim 1.6$ for HSs) we expect anyway that asymmetric HS-HS systems which collapse promptly to a BH would leave tori significantly lighter that the ones produced within the one family scenario (for which $\mtov$ must be larger than $2\ms$.)}
\vskip 0.2cm
2) $\mshyb /(1-\alpha) <\mtot < \mth$. Within the parameters' space
discussed above such a possibility is never realized.

\vskip 0.2cm
3) $\mtot<\min [\mshyb/(1-\alpha),\mth]=\mth$: 
the postmerger remnant is initially a hypermassive HS star.
The conversion of hadronic matter into quark matter takes place as described in the previous sections and it is responsible for a dramatic change of the structure of the star: differential rotation can be enhanced by the presence of two distinct phases and the radius increases \citep{Drago:2015fpa}. The spectrum of the GW's signal associated with the oscillations of the remnant, initially peaked at very high frequencies, should progressively shift towards smaller frequencies while quark matter is produced \citep{Bauswein:2015vxa}. The remnant transforms into a QS (on a time scale of seconds) and is either a supramassive or (if $\mtov^Q$ is very large) a stable star.
An extended X-ray emission following the sGRB can take place and a KN signal is expected.

\subsubsection{HS-QS merger.}

1) $\mtot > \mthyb$: this case is similar to the analogous case of HS-HS merger. A possible difference could come from the inspiral GW's signal since we expect a larger value of $\tilde{\Lambda}$ with respect to the case of HS-HS systems.
\vskip 0.2cm
2) $ \mshyb/(1-\alpha) < \mtot<\mthyb$: the remnant is a hypermassive star.
Within the two-families scenario this is the only possible interpretation of the event of August 2017. We will discuss that event separately, in Sec.\ref{gw17}.

3) $\mtot< \mshyb/(1-\alpha)$: the phenomenology is similar to case (3) of HS-HS merger and it is characterized by the possibility of producing an extended emission following the sGRB.

\subsubsection{QS-QS merger.}

1) $\mtot>\mtq\sim\msup^Q/(1-\alpha)$: this case is most probably irrelevant from the phenomenological point of view since the mass distribution of binary systems  peaks at $1.33\ms$ with a $\sigma \sim 0.11 \ms$ \citep{Kiziltan:2013oja} and the value of $\msup^Q\sim 3\ms$.
Its phenomenology would anyway be very similar to the ones of the previously discussed cases of prompt collapse.
\vskip 0.2cm

2) $\mtot<\mtq\sim\msup^Q/(1-\alpha)$.
This case was explored long time ago by \citet{Haensel:1991um} as a possible mechanism for producing sGRB.
In particular, it was recognized that the environment around the merger is relatively clean from baryon pollution and therefore it is easier to produce ejecta with a high Lorentz factor.
The ultimate source of the sGRB in their analysis is provided by the cooling of the remnant and the spectra would therefore be very similar to that of a black body. Notice however that rotation and magnetic field were not considered in that pioneering study. 
If the post-merger object rotates rapidly and it develops a strong magnetic field, then the protomagnetar mechanism could be applied also to this merger, and an extended emission could also be generated as in the other cases discussed above.
A completely open issue concerns the possible
presence and the features of a KN associated with such an event. In a recent study of \citet{Paulucci:2017opy}, it has been proposed that the evaporation of strangelets could be very efficient and that r-process
nucleosynthesis can be at work although not for the production
of lanthanides. The KN associated with that event could thus show a dominant blue component, but further and more detailed studies are necessary, since the chain of nucleosynthesis could possibly be completed also in this case.

\begin{figure*}%[h!]
	\begin{centering}
    	\epsfig{file=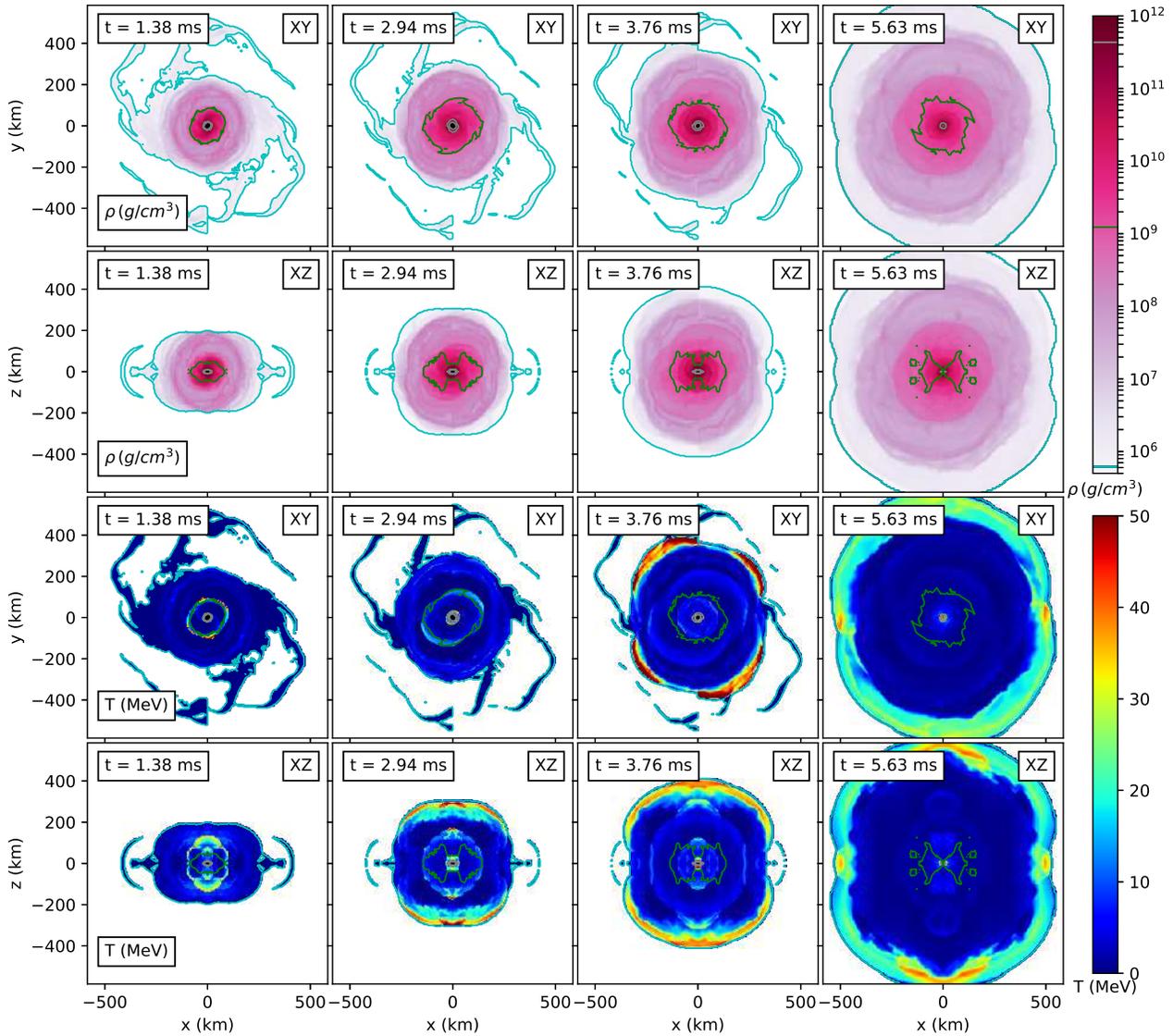,width=17cm}
        %\vspace{-0.5cm}
        \caption{Snapshots of the XY and XZ projections of the time component of  the density, $\rho$, and the temperature, $T$, for the SFHo-HD 118vs118 model. In each panel, the white region outside the solid cyan line refers to the atmosphere, which was excluded from the computation, while the colored lines refer to the density contour which correspond to the densities used for the polytropic approximation discussed in Section \ref{sec:numeric}. 
       \label{fig:RHOT} 
        }
    \end{centering}
    %\pagebreak
\end{figure*}

\begin{figure*}%[h!]
	\begin{centering}
    	\epsfig{file=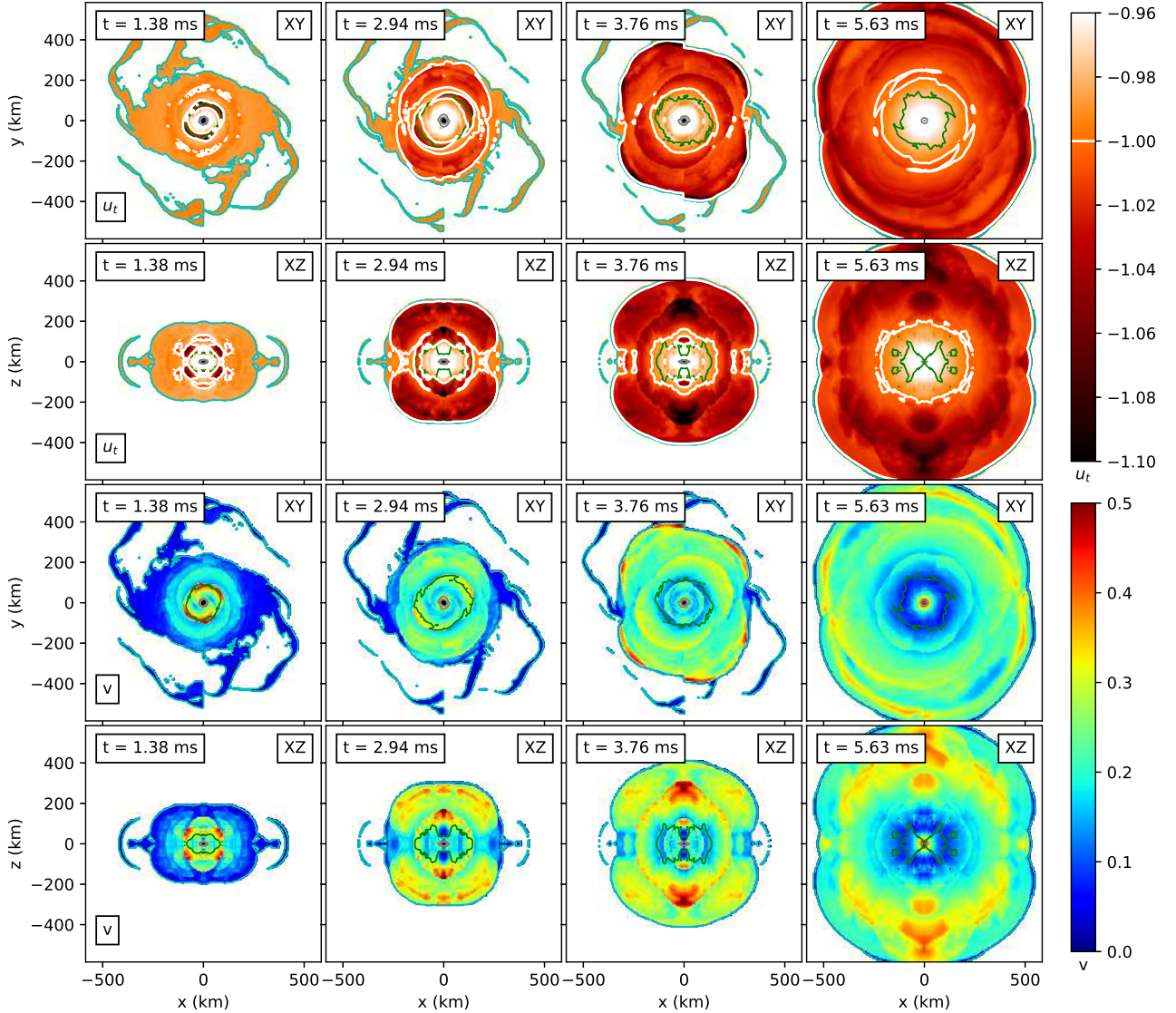,width=17cm}
        %\vspace{-0.5cm}
        \caption{Snapshots of the XY and XZ projections of the time component of the fluid velocity, $u_t$, and the matter velocity, $v$, for the SFHo-HD 118vs118 model. In each panel, the white region outside the solid cyan line refers to the atmosphere, which was excluded from the computation, while the colored lines refer to the density contour which correspond to the densities used for the polytropic approximation discussed in Section \ref{sec:numeric}. 
        The solid white line in the first two rows refers to the $u_t=-1$ condition, which defines the difference from bound and unbound matter, as discussed in section \ref{sec:ejected}. The times $t$ is relative to the merger time for this model.
\label{fig:vel} 
        }
    \end{centering}
    %\pagebreak
\end{figure*}

\begin{table}
\begin{center}
\begin{tabular}{|c||c|c|c|c||c|c|c|}
\hline
EOS & $M$         & $R$ & $C$ & $\Lambda$ & $M_{ADM}$  & $J_{ADM}$ & $\Omega_0$ \\
    & ($M_\odot$) & (km)&     &           &($M_\odot$) & (CU)      & {\tiny (krad/s)}   \\
\hline \hline
SFHo-HD & {1.18} & 11.166 & 0.156 & 544 & 2.340 & 5.783 & 1.713 \\
\hline
SFHo-HD & {1.20} & 11.133 & 0.159 & 483 & 2.379 & 5.945 & 1.725 \\
\hline
SFHo-HD & {1.22} & 11.100 & 0.162 & 429 &2.419 & 6.092 & 1.736 \\
%\hline
%%SFHo-HD & {\bf 1.22}(1.24) & 11.253 & 0.148 & 2.431 & 6.174 & 1.741 \\
%%SFHo-HD & \multicolumn{4}{c||}{{\bf 1.22 - 1.24 } (unequal mass)} & 2.431 & 6.174 & 1.741 \\
\hline
SFHo-HD & {1.24} & 11.067 & 0.165 & 382 & 2.451 & 6.257 & 1.747 \\
\hline
SFHo-HD & {1.26} & 11.034 & 0.169 & 340 & 2.490 & 6.427 & 1.759 \\
\hline
SFHo-HD & {1.28} & 11.002 & 0.171 & 303 & 2.530 & 6.597 & 1.770 \\
\hline
SFHo-HD & {1.30} & 11.970 & 0.175 & 270 & 2.570 & 6.771 & 1.781 \\
%%%%%%%%%%%%%%%%%%%%%%%
%[[ 1.18       11.16597369  0.15604246]
% [ 1.2        11.13258155  0.15916323]
% [ 1.22       11.09958183  0.16229704]
% [ 1.24       11.06689403  0.16544487]
% [ 1.26       11.03440426  0.16860833]
% [ 1.28       11.00213827  0.17178698]
% [ 1.3        10.96968085  0.17498738]]
\hline\hline
SFHo & 1.18 & 11.942 & 0.145 & 944 & 2.332 & 5.765 & 1.714 \\ \hline
SFHo & 1.20 & 11.939 & 0.148 & 856 & 2.372 & 5.926 & 1.725 \\ \hline
SFHo & 1.22 & 11.934 & 0.151 & 777 & 2.431 & 6.174 & 1.742 \\ \hline
SFHo & 1.24 & 11.929 & 0.153 & 705 & 2.450 & 6.258 & 1.748 \\ \hline
SFHo & 1.26 & 11.925 & 0.156 & 641 & 2.490 & 6.426 & 1.759 \\ \hline
SFHo & 1.28 & 11.919 & 0.159 & 582 & 2.529 & 6.597 & 1.770 \\ \hline
SFHo & 1.30 & 11.913 & 0.161 & 530 & 2.569 & 6.770 & 1.781 \\ \hline
SFHo & 1.32 & 11.907 & 0.164 & 482 & 2.608 & 6.945 & 1.791 \\ \hline
SFHo & 1.34 & 11.900 & 0.166 & 439 & 2.647 & 7.121 & 1.801 \\ \hline
SFHo & 1.36 & 11.893 & 0.169 & 399 & 2.687 & 7.299 & 1.812 \\ \hline
SFHo & 1.38 & 11.885 & 0.171 & 364 & 2.726 & 7.480 & 1.822 \\ \hline
SFHo & 1.40 & 11.877 & 0.174 & 332 & 2.766 & 7.663 & 1.832 \\ \hline
SFHo & 1.42 & 11.868 & 0.177 & 302 & 2.805 & 7.847 & 1.843 \\ \hline
\end{tabular}
\end{center}
\caption{Main properties of the equal mass models considered in this work, labeled by the name of the EOS. The table include the gravitational mass, $M$, the radius, $R$, and the compactness, $C$, of the individual stars. In addition, here it is reported the total ADM mass, $M_{ADM}$, and angular momentum, $J_{ADM}$, and the angular velocity, $\Omega_0$, at the beginning of the simulations. All the models have an initial separation of roughly 44.3 km.
\label{tab:PROP}
}
\end{table}

\section{Numerical simulations of the merger of HS-HS}\label{sec:numeric}

We considered Binary Neutron Star merger processes using numerical relativity in the 
case of purely HSs and for two EoSs, namely the SFHo 
and SFHo-HD EOS, presented in Section \ref{sec:EOS} and we studied the behavior of equal mass 
models, as summarized in Table \ref{tab:PROP}, for both of them.

\subsection{Numerical methods and initial data}
The numerical methods used to perform these simulations is the same of \citet{DePietri2018,DePietri2015,Maione:2017aux,Maione2016}. We report here the general simulation setup and parameters and refer to those previous articles for more details. In particular, the resolution used in this work is $dx=0.1875$ CU $= 277$ m (see \citet{DePietri2015} for a discussion of the convergence properties of the code).

The simulations were performed using the \codename{Einstein Toolkit} \citep{Loffler:2011ay}, an open source, modular code for numerical relativity based on \codename{Cactus} \citep{CactusUsersGuide:web}. The evolved variables were discretized on a Cartesian grid with 6 levels of fixed mesh refinement, each using twice the resolution of its parent level. The outermost face of the grid was set at $720M_\odot$ ($1040$ km) from the center. We solved the BSSN-NOK formulation of the Einstein's equations \citep{BSSN:1987Nakamura,BSSN:1995Shibata,BSSN:1998Baumgarte,BSSN:2000Alcubierre,BSSN:2003Alcubierre} implemented in the \codename{McLachlan} module \citep{Brown:2008sb}, and the general relativistic hydrodynamics equations with \textit{high resolution shock capturing methods}, implemented in the publicly available module \codename{GRHydro} \citep{GRHydro:2005Baiotti,GRHydro:2014code}. In particular, we used a finite-volume algorithm with HLLE Riemann solver \citep{HLL,HLLE} and the WENO reconstruction method \citep{WENO:1994jcp,WENO:1996jcp}. The combination of the BSSN-NOK formalism for the Einstein's equations and the WENO reconstruction method was found in \citet{DePietri2015} to be the best setup within the \codename{Einstein Toolkit} even at low resolutions. For the time evolution we used the method of lines, with fourth-order Runge-Kutta \citep{Runge1895,Kutta:1901aa}. 
For numerical reasons the system is evolved on an external matter atmosphere set to   
$\rho_{atm}= 6.1 \cdot 10^5$ g/cm${}^3$  (as in  \cite {Lehner:2016lxy}) that it is just slightly larger than the one used in \cite{Shibata:2015prd} and \cite{Radice:2018pdn} but it is still sufficiently low to avoid atmosphere's 
inertial effects on the ejected mass on a time scale of  $\simeq 10$ ms after the  onset of the merger
\cite{Shibata:2015prd}.  Initial data were generated with the \codename{LORENE} code \citep{Lorene:2001Gourg}, as irrotational binaries in the conformal thin sandwich approximation. In this work we analyze different combinations of equal mass setups for binary systems using the SFHo-HD \citep{Drago:2014oja,Burgio:2018yix} and the SFHo EOSs \citep{Steiner:2012rk}. For both models, the initial distance was set to $44.3$ km like in \citet{Maione2016,Feo:2016cbs}. Matter description is performed using  
a ten piece piece-wise polytropic approximation for an EoS of the type $P = P(\rho,\epsilon)$ supplemented by an additional thermal component described by $\Gamma_{th}=1.8$. The parametrization of the EoS uses the following prescriptions:
\begin{equation}
\epsilon = \epsilon_0(\rho) + \epsilon_{th}
\end{equation}
\begin{equation}
p = p_0(\rho) + (\Gamma_{th}-1)\rho \epsilon_{th}
\end{equation}
where $\epsilon_{th}$ is an arbitrary function of the thermodynamical state that has the property of being zero at $T=0$ and $\epsilon_0 (\rho)$ and $p_0 (\rho)$ are the internal energy and internal pressure at $T=0$. In particular, for a piece-wise polytropic approximation with $N$ pieces we can write
\begin{equation}
p_0 (\rho) = K_i \rho^{\Gamma_i}
\end{equation}
\begin{equation}
\epsilon_0 (\rho) = \epsilon_i + \frac{K_i}{\Gamma_i -1} \rho^{\Gamma_i -1}
\end{equation}
and all the coefficients are set once the polytropic index $\Gamma_i$ ($i=0$, ... , $N-1$), the transition density $\rho_i$ ($i=1$, ... , $N-1$) and $K_0$ are chosen. One should note that $\epsilon_{th}$ here plays the role of temperature and it may be converted (using its interpretation as a perfect fluid thermal component) to a temperature scale as $T=(\Gamma_{th}-1) m_{b} \epsilon_{th}$ where we assumed for the mass of a free baryon a conventional value of $m_{b}=940$ MeV/$c^2$.

\subsection{Gravitational waves extraction}\label{subsec:GW}

During the simulations, the GW signal is extracted using the Newman-Penrose scalar $\Psi_4$ \citep{PSI4:1962jmp,PSI4:2002prd} using the module \codename{WeylScalar4}. The scalar is linked to the GW strain by the following relation, which is valid only at spatial infinity:
\begin{equation}
\Psi_4 = \ddot{h}_{+} - i\ddot{h}_{\times}
\end{equation}
where $h_+$ and $h_{\times}$ are the two polarization of the GW strain $h$. The signal is then decomposed into multipoles using spin-weighted spherical harmonics of weight $(-2)$ \citep{SpherHarm:1980Thorne}. This procedure is made by the module \codename{MULTIPOLE} using the following relation
\begin{equation}
\psi_4(t,r,\theta,\phi) = \sum_{l=2}^{\infty} \sum_{m=-l}^{l} \psi_4^{lm}(t,r) _{-2}Y_{lm}(\theta, \phi) \, .
\end{equation}

In this work we focus only on the dominant $l=m=2$ mode, therefore we will refer to $h_{2,2}$ as $h$ for the rest of this paper. In order to extract the GW strain form $\Psi_4$ and minimize the errors due to the extraction, one has to extrapolate the signal extracted within the simulation at finite distance from the source to infinity, in order to satisfy the previous equation. Then the extrapolated $\Psi_4$ is integrated twice in time, employing an appropriate technique in order to reduce the amplitude oscillations caused by the high-frequency noise aliased in the low-frequency signal and amplified by the integration process \citep{Integ:2011cqg,Integ:2015prd}. The procedure adopted in this work is extensively discussed in \citep{Maione2016}: first $\Psi_4$ is extrapolated to spatial infinity using the second order perturbative correction of \citep{Integ:2015prd},
\begin{align}
 R\psi_4^{lm} (t_{ret}) |_{r=\infty} = &\bigg( 1-\frac{2M}{R} \bigg) \bigg( r\ddot{\bar{h}}^{lm} (t_{ret}) \\
 &-\frac{(l-1)(l+2)}{2R}\dot{\bar{h}}^{lm} (t_{ret}) \nonumber\\
 &+\frac{(l-1)(l+2)(l^2+l-4)}{8R^2} \bar{h}^{lm} (t_{ret}) \bigg),
 \nonumber
\end{align}
where the GW strains ($\bar{h}^{lm} $) at finite radius 
(in our case at a fixed coordinate radius $R=1033$ km) is computed by 
integrating the Newman-Penrose scalar twice in time with a simple trapezoid rule, starting from zero coordinate time, and fixing only the two physically meaningful integration constants $Q_0$ and $Q_1$ by subtracting a linear fit of itself from the signal,
\begin{align}
    \bar{h}_{lm}^{(0)} = \int_0^t dt' \int_0^{t'} dt'' \psi_4^{lm} (t'',r),
\\
\bar{h}_{lm} = \bar{h}_{lm}^{(0)} - Q_1 t - Q_0
\end{align}
After the integration, we apply a digital high-pass Butterworth filter, which is designed to have a maximum amplitude reduction of $0.01$ dB at the initial GW frequency $f_{t_0}$ (assumed to be 2 times the initial orbital angular velocity, as reported by the \codename{LORENE} code) and an amplitude reduction of $80$ dB at frequency $0.1f_{t_0}$. All the GW related information is reported in a function of the retarded time
\begin{align}
&t_{ret} = t - R^* \\
&R^* = R+2M_{ADM} \log{\left( \frac{R}{2M_{ADM}}-1 \right)} 
\end{align}
The physical observable we are focusing on in this work, which is the GW amplitude spectral density $|\tilde{h}(f)|f^{1/2}$, is calculated from the GW strain obtained with the aforementioned procedure using
\begin{equation}
    |\tilde{h}(f)| = \sqrt{\frac{|\tilde{h}_+(f)|^2 + |\tilde{h}_\times|^2}{2}},
\end{equation}
where $\tilde{h}(f)$ is the Fourier transform of the complex GW strain,
\begin{equation}
\tilde{h}(f) = \int_{t_i}^{t_f} h(t) e^{-2\pi i f t} dt
\end{equation}
and the associated GW energy in the post merger phase in the interval that span 
from $1$ ms up to $15$ ms after the merger time. On this respect, we have defined the
merger time as the time at which the $l=2,m=2$ GW strain ${h}^{lm}$ has a maximum. 
All times are reported by fixing $t=0$ as the merger time.  

\begin{figure}
    \centering
    \epsfig{file=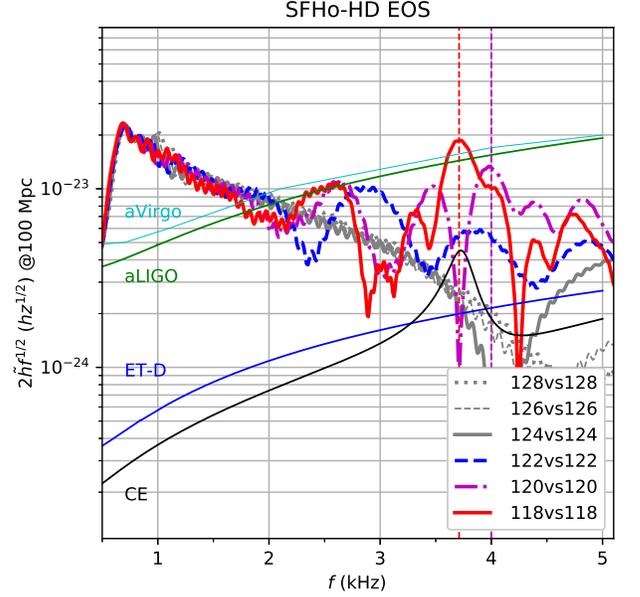,width=8.5cm}
    \\[-2mm]
    \caption{Spectrum of the GW signal for models described by the SFHo-HD EoS. All the given models show a reduced $f_2$ post merger peak that is
    fully suppressed for model that have a direct collapse to a Black-Hole ($\mtot\geq 2.48 \ms$.}
    \label{fig:SpectrumDP}
\end{figure}

\begin{figure}
    \centering
    \epsfig{file=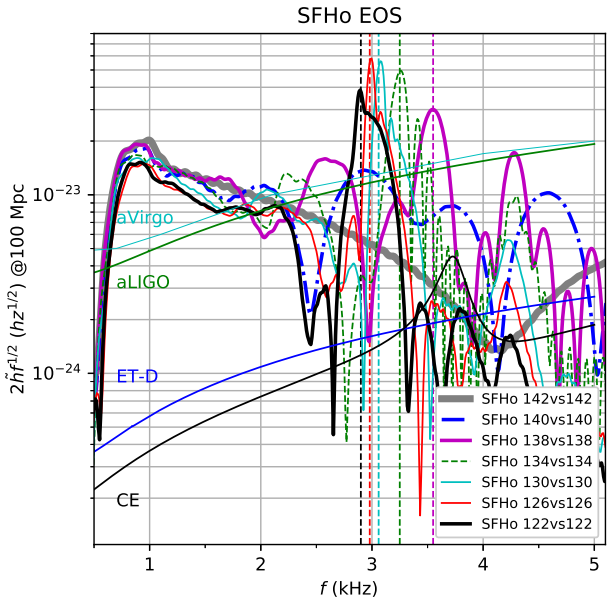,width=8.5cm}
    \\[-2mm]
    \caption{Spectrum of the GW signal for models described by the SFHo EoS. Note that the model with total mass 2.84 $M_\odot$ has no post merger peak. The model with total mass 2.80 $M_\odot$ show just a marginal main $f_2$ peak while all the less massive model show a distinct main $f_2$ post merger peak and two (of lower amplitudes) side peaks (forming a three peaks structure) that become less import as the total mass decrease. }  
    \label{fig:SpectrumSFHo}
\end{figure}

\subsection{Estimate of the ejected mass and disk mass}\label{sec:ejected}

The estimate of the ejected mass has been performed using the following condition to determine if the matter is gravitationally bound or not. 
We look at the time component of the fluid four-velocity $u_t$ and we use the local condition 
$u_t < -1$ that provides a necessary although not sufficient condition for a fluid element to 
be unbound \citep{Rezzolla2010}. 
We have indeed computed  the total unbound mass present on the computational domain (using the above criteria $u_t < -1$, and not considering the part of the computational grid that correspond to the atmosphere) at each simulation time and assuming as a measure of the total ejected mass the maximum of such values observed in the whole simulation time, that we denoted as $M_{\mathrm{ej}}$. 
At the same time we have estimated the disk-mass $M_{\mathrm{disk}}$ (reported in Table \ref{tab:PROP}) as the total mass present in the computational grid having a density $\rho \leq 3.3 \cdot 10^{11}$ g/cm$^{3}$ and that it is bound (i.e., having $u_t > -1$)  20 ms after merger. Such a way of estimating $M_{\mathrm{disk}}$ allows to make a meaningful quantification of the mass surrounding the star that is not yet unbound due to dynamical ejection.

Another possibility to estimate the total ejected mass is to compute
the flow (we used the module \codename{Outflow}) of the material leaving a  coordinate sphere surface with a given radius using the $u_t < -1$ condition.
Using this procedure we determined the unbound matter flow for a set of different radii from $65$ CU $\simeq 96$ km  up to $700$ CU $\simeq 1034$ km, which is the border of the computational domain.

\subsection{Results of the numerical simulations}

In this work we simulate equal mass binary systems for HS-HS and NS-NS configurations during about four orbits before merger and for about 20 ms after merger.

\begin{figure}
	\begin{centering}
		\epsfig{file=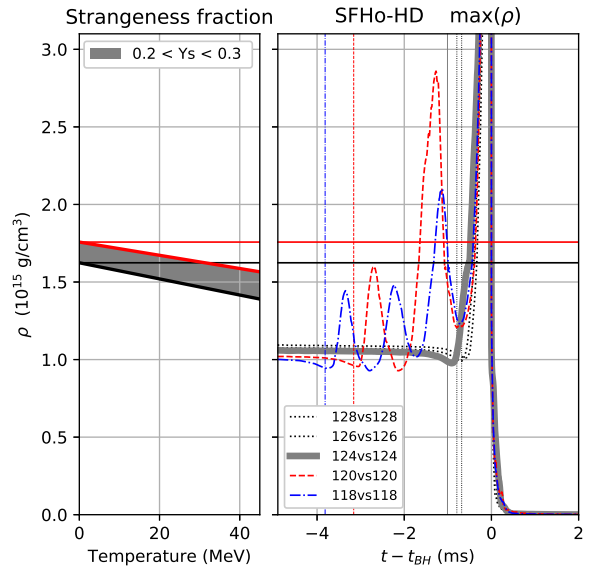,height=8cm,width=8cm}
       \\[-4mm]
		\caption{On the left it is shown the density for which the strangeness fraction is 0.2 (Black line) 
		and 0.3 (Red line) as a function of the temperature in MeV. On the right, it  is shown the maximum mass density as function of time for the models described by the SFHo-HD EoS.   
        \label{fig:BounceDP}
        }
	\end{centering}
\end{figure}

\begin{figure}
	\begin{centering}
		\epsfig{file=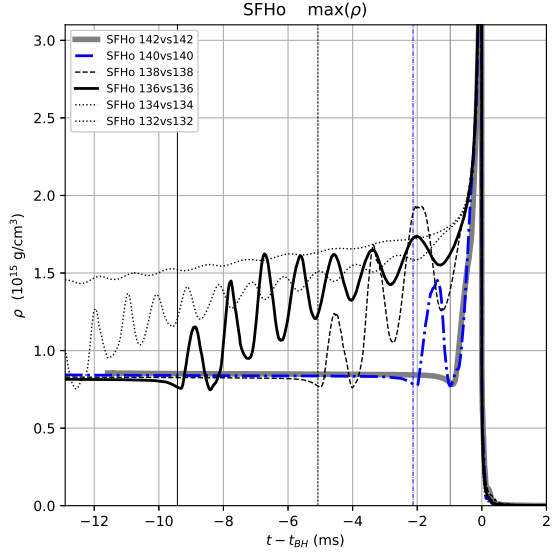,height=8cm,width=8cm}
       \\[-4mm]
		\caption{Plot of the maximum density as function of time for models described by the SFHo EoS. It should be noted that the first model showing no bouncing corresponds to a total mass of 2.84 $M_\odot$.     
        \label{fig:BounceSFHo}
        }
	\end{centering}
\end{figure}

All the simulated model show a very similar behaviour. To show the main properties of the evolution snapshots for model SFHo-HD 118vs118 are shown in Fig.~\ref{fig:RHOT} and in Fig.~\ref{fig:vel}. In particular in Fig.~\ref{fig:RHOT} are shown the density of the expelled matter and its temperature at different times, while in Fig.~\ref{fig:vel} its velocity and the localization of unbound mass at the same times.  
The computed GW signal of our simulation is shown in Fig. \ref{fig:SpectrumDP} and  Fig. \ref{fig:SpectrumSFHo},
where we display the power spectrum extracted using a detector at $R=1034$ km for models 
described by the SFHo-HD EoS and the SFHo EoS, respectively. Associated to each simulation we report the values of $M_{\mathrm{ej}}$ and of $M_{\mathrm{disk}}$ in Table \ref{tab:MejMdisk}. Their values are also plotted in Fig. \ref{fig:EjectedMass} as a function of $\mtot$. A striking feature of the GW spectrum is that models with 
$\mtot  \geq 2.48$ and $\mtot \geq 2.84$, for SFHo-HD and SFHo respectively, do not show any post merger $f_2$ peak. This is related to the fact that, in these cases, we have a direct collapse to BH (less than 1 ms after the merger). The fact that a prompt collapse occurs is clearly shown in the plot of the maximum density as a function of time for models described by SFHo-HD EoS (Fig. \ref{fig:BounceDP}) and by the SFHo EoS (Fig. \ref{fig:BounceSFHo}). In the cases in which a prompt collapse does not occur, the main structure of the spectrum is the same as the one discussed in \citet{Takami:2014tva,bauswein:2015unified,Maione:2017aux} and references therein. Importantly, the post merger spectrum is characterized by a main $f_2$ peak whose frequency depends on the EoS. Moreover, its frequency increases with the mass of the star while its tidal parameter decreases. One should also notice that secondary peaks are present but their relative importance decreases as the total mass of the system is reduced. 


The presence of a well defined main peak and its frequency $f_2$ could lead to a clear determination of the kind of merger and of the properties of the EoS. In fact, that would be a precise signature that the system did not directly collapse to a BH and its spectrum encodes
information on the properties (mainly the stiffness) of dense matter.
For example, the merger of two compact stars of mass 1.18 $M_\odot$ shows a main peak at frequency $f_2=3.71$ kHz or $f_2=2.88$ kHz if the EoS describing matter is SFHo-HD or SFHo, respectively. Unfortunately the amount of GW energy available for the mode detection (see Table \ref{tab:MejMdisk}) is going to be at best of the order of 0.1 $M_\odot$ and therefore it is unlikely that such a  peak will be detected by the present generation of GW detectors but they will be a distinct feature to be observed in third generation detectors.

%%%%%%%%%%%%%%%%%%%%%%%%
%%%
%%%%%%%%%%%%%%%%%%%%%%%%

\begin{figure}
	\begin{centering}
		\epsfig{file=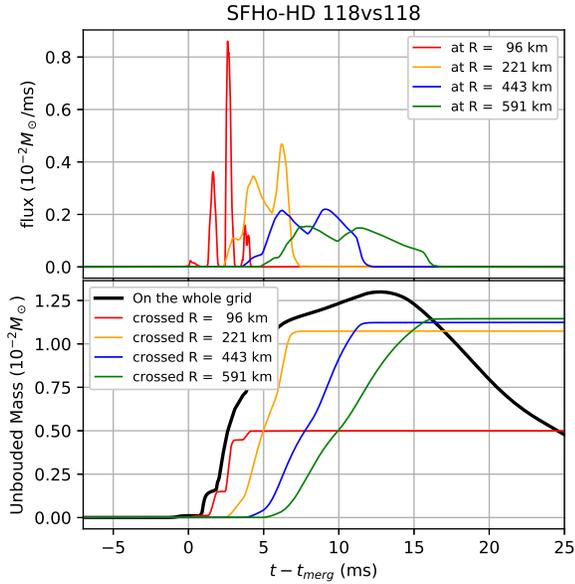,height=8cm,width=8cm}
		\caption{Upper panel: estimates of the unbound material by using the $u_t <-1$ condition, from the flow crossing spherical surfaces with different coordinate radii. We show the results at four radii.
		We note that there are three peaks at the closest surface, which correspond to the "rebounds" of the hypermassive neutron star formed after the merger. Those peaks become larger when crossing the farther surfaces due to the different velocities of the ejected material. Lower panel: time integrated fluxes at different crossing radii. The black line corresponds to the total unbound mass on the whole grid at a given time.
    \label{outflow} 
    }
	\end{centering}
\end{figure}

\begin{figure}
	\begin{centering}
		\epsfig{file=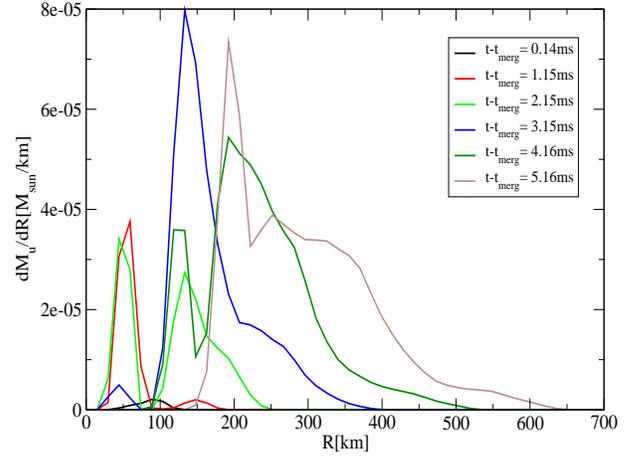,height=8cm,width=6cm,angle=-90}
		\caption{Snapshots at different times of the radial density of the unbound matter as a function of the radius for the SFHo-HD 118vs118 model. The appearance of various peaks at a given radius and for different times corresponds to the shocks generated by the oscillations of the remnant displayed in Fig.8.
        \label{fig:dmdr} 
        }
	\end{centering}
\end{figure}

Let us discuss now the results for the ejected mass and the amount of mass which is left into the disk.
In Fig.\ref{outflow} we show an example of the outflow for the SFHo-HD 118vs118. By following the flow of the unbound matter during the evolution,  we can estimate the ejected mass by integrating (at each radius) on time the unbound matter flow. In this model, the total flows of unbound mass at coordinate radius 96, 220, 443 and 590 km are: (5.0, 10.7, 11.24, 11.45) $\mathrm{m}M_\odot$, respectively. Indeed, this procedure for computing the total ejected mass is very sensitive to the extraction radius and this is a common property of all the simulated models.
In particular we compared the results obtained by using this method with those obtained by considering the maximum of total unbound mass present on the whole computational domain at a given time. We note that the difference between 
these two estimates of $M_{\mathrm{ej}}$ is always less than about 20\% and the two results get closer by increasing the extraction radius.
From the lower panel of Fig.~\ref{outflow} one should note that most of the dynamically unbound mass is generated between 100 km and 200 km from the center of the remnant and it is related to three main outbursts of ejected matter associated to the three bouncing of the maximum density. This is more clear looking at Fig.~\ref{fig:dmdr}  where we display the radial density profile (as a function of the coordinate radius) of the unbound matter at different times. There, one should note that three shocks are created within a 100 km radius from the center of the star ($t$=0.14ms, $t$=1.15ms and $t$=2.15ms) the first being of lower amplitude. Then they move out, amplify and spread while the system evolves. This is analogous to what is observed for the flux of matter at different radii (see Fig. \ref{outflow}, upper panel) where one can see that at $R=100$ km one has three clear separated peaks that join and spread as the unbound matter flows outside the computational domain. 

The same overall dynamics is associated to the mass ejection for all the considered models (except models that feature a direct collapse to BH and for which the mass ejection is suppressed). All the results for $M_{\mathrm{ej}}$ and $M_{\mathrm{disk}}$ as a function of $\mtot$ are summarized in the lower and upper panel, respectively, of Fig. \ref{fig:EjectedMass}. The explicit values are also reported in Table \ref{tab:MejMdisk}. 
A striking feature concerning 
the mass ejected is the presence of a maximum which is located 
at a value of $\mtot$ slightly smaller 
than the value of $M_{\mathrm{threshold}}$. This is one of the main results of the present work.
In particular this maximum is located at $2.72 \ms$ for the SFHo EoS and it corresponds 
to an ejected mass of $\simeq 16$ m$M_\odot$. For the SFHo-HD case,
we have not determined the maximum of the ejected mass
which should anyway be located below or close to $\sim 2.36 \ms$ (the simulation with the lowest $\mtot$). For this specific value we have found an ejected mass of $\simeq 13$m$M_\odot$, very close to the maximum for SFHo EoS.
The plots of the ejected mass show a very steep decay of its value as the total mass of the binary increases and the ejection is almost completely suppressed in the case of a prompt collapse to BH.

%%%%%%%%%%%%%%%%%%%%%%%%%%%%%%%%%%%%%%%%%%%%%%%%%%%%%%%%%%
%%% Eject and disk masses
%%%%%%%%%%%%%%%%%%%%%%%%%%%%%%%%%%%%%%%%%%%%%%%%%%%%%%%%%%

\begin{table}
\centering
\begin{tabular}{| c | r | r | r | c | r | }
\hline
Model & $M_{\mathrm{ej}}$~~~  & $M_{\mathrm{disk}}$~~~  & $E_{gw}^{POST}$~  & $f_2$ & $t_{BH}$ \\[1mm]
      & (m$M_{\odot}$) 
      & (m$M_{\odot}$)
      & (m$M_{\odot}$)
      & (kHz)
      & (ms) \\
\hline \hline
SFHo-HD 118vs118 &   12.993 &  12.92  &  25.42    &  3.71  & 3.82 \\ %\hline
SFHo-HD 120vs120 &    9.435 &  13.81  &  22.42    &  4.00  & 3.16\\ %\hline
SFHo-HD 122vs122 &    4.290 &   8.34  &   6.06    &  ----  & 1.91\\ %\hline
SFHo-HD 124vs124 &    3.011 &   2.89  &   0.66    &  ----  & 1.00\\ %\hline
SFHo-HD 126vs126 &    0.737 &   2.45  &   0.20    &  ----  & 0.79\\ %\hline
SFHo-HD 128vs128 &    0.055 &   0.74  &   0.04    &  ----  & 0.70\\ %\hline
SFHo-HD 130vs130 &    0.043 &   0.71  &   0.01    &  ----  & 0.59\\ %\hline
\hline\hline
SFHo 118vs118    &    1.968 &  76.66  &  42.16    &  2.88  & - - - \\ %\hline
SFHo 120vs120    &    2.085 &  71.72  &  43.87    &  2.90  & - - - \\ %\hline
SFHo 122vs122    &    1.730 &  91.81  &  42.00    &  2.90  & - - - \\ %\hline
SFHo 124vs124    &    1.824 &  65.58  &  52.98    &  2.96  & - - - \\ %\hline
SFHo 126vs126    &    2.375 &  60.86  &  58.33    &  2.98  & - - - \\ %\hline
SFHo 128vs128    &    3.145 & 112.24  &  50.33    &  3.05  & - - - \\ %\hline
SFHo 130vs130    &    4.523 &  73.82  &  59.33    &  3.06  & - - - \\ %\hline
SFHo 132vs132    &    6.007 &  88.87  &  67.29    &  3.18  & 25.75 \\ %\hline
SFHo 134vs134    &    9.511 &  49.27  &  65.09    &  3.25  & 13.55 \\ %\hline
SFHo 136vs136    &   16.244 &  30.71  &  58.76    &  3.40  & 9.42  \\ %\hline
SFHo 138vs138    &   10.367 &  16.09  &  46.06    &  3.55  & 5.06  \\ %\hline
SFHo 140vs140    &    4.170 &   6.45  &  22.39    &  ----  & 2.13  \\ %\hline
SFHo 142vs142    &    2.247 &   2.01  &   2.02    &  ----  & 0.98  \\
\hline
\end{tabular}
\caption{Values of  ejected mass ($M_{\mathrm{ej}}$), disk mass  ($M_{\mathrm{disk}}$), energy emitted in GW in the postmerger phase ($E_{gw}^{POST}$), frequency of the main peak of GW signal  ($f_2$),  and time to collapse to BH ($t_{BH}$) for all the models
considered in this work. 
        \label{tab:MejMdisk} 
        }
\end{table}

\begin{figure}
	\begin{centering}
    	\epsfig{file=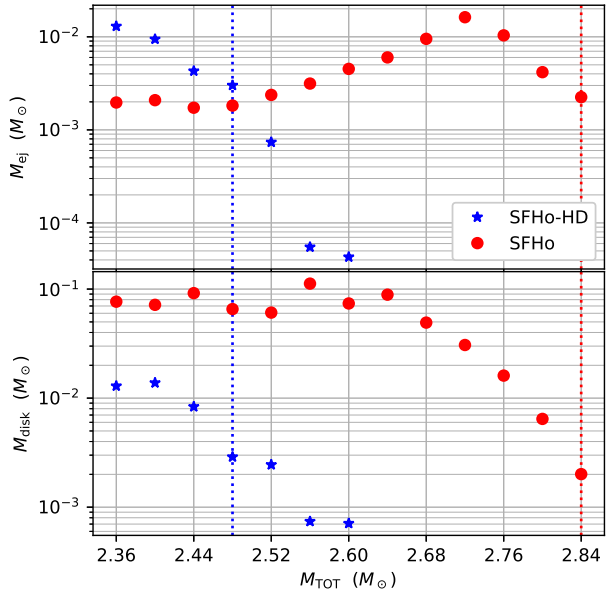,height=8.5cm,width=8.5cm}
        \caption{Representation of the ejected mass and disk mass as reported in 
        TABLE ~\ref{tab:MejMdisk} in terms of the total mass of the model. The black dots
        refer to the SFHo-HD models, while the red dots refer to the SFHo ones.
        The dotted vertical lines correspond to the values of $M_{\mathrm{threshold}}$ for both 
        equations of state.}
        \label{fig:EjectedMass} 
    \end{centering}
\end{figure}

Another clear difference between 
the two EoSs concerns the value of $M_{\mathrm{disk}}$:
in the case of SFHo-HD it
is always $\lesssim 0.01\ms $ whereas
for SFHo it can be an order of magnitude larger. 
These are potentially very interesting 
results for the phenomenology of the KN. As we will discuss in Sec. \ref{sec:conclusions}, the luminosity of the different components of the KN (red, blue and purple \citep{Perego:2017wtu}) strongly depends on the specific mechanisms for the ejection of mass during the different stages of the merger. A relevant fraction of the ejected mass can in principle come from the disk and therefore a very small value of $M_{\mathrm{disk}}$ corresponds to a strong suppression of certain components of the KN signal.

\begin{figure}
	\begin{centering}
		\epsfig{file=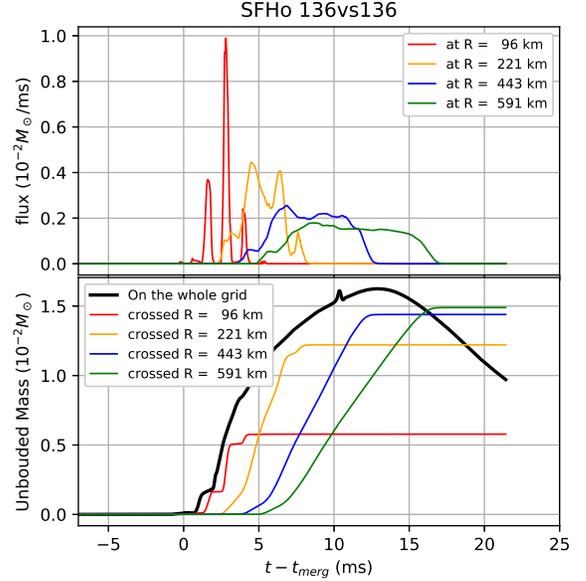,height=8cm,width=8cm}
		\caption{Figure of the material unbound, using the $u_t <-1$ condition, crossing coordinate spherical surfaces with different radii. In this figure, we report the outcome of four of them for the SFHo model with mass of $1.36$ $M_\odot$, in particular the ones corresponding to a radius of $65$ CU $\simeq 96$ km, $150$ CU $\simeq 221$ km, $300$ CU $\simeq 443$ km and $400$ CU $\simeq 591$ km. We note that also for this models there are three peaks at the closest surface, which correspond to the "rebounds" of the hypermassive neutron star formed after the merger. Those peaks become larger when crossing the farther surfaces due to the different velocities of the ejected material. Showing that these feature are common to the case of both the two EoSs considered here. 
    \label{outflowSFHo} 
    }
	\end{centering}
	\vspace{0.1cm}
\end{figure}

Let us now compare our results with the ones  obtained by similar numerical simulations present in the literature. In particular there are a few simulations using the SFHo EoS for masses similar to our model SFHo 136vs136.
 The outflow dynamics for this model is shown in Fig. \ref{outflowSFHo}.  The results we obtain are in general agreement with results obtained in \citet{Shibata:2015prd,Shibata:2016prd} for the SFHo EoS. In those works it was found that the associated ejected mass is of the order of 10 m$M_\odot$ for the mass of the two stars of 1.35 $M_\odot$ and of the same order of magnitude for unequal mass systems of approximately the same total mass \citep{Shibata:2017xdx}. In the same work, it was also found that the expected disk mass (for the same systems) is of the order of 50 m$M_\odot$ up to 120 m$M_\odot$ and indeed of the same order of magnitude of the values found in the present work. Computation of the ejected mass for the SFHo EoS were also presented in \citet{Bauswein:2013yna} using SPH dynamics, in \cite{Lehner:2016lxy} and, using WhiskyTHC code,  in \cite{Bovard:2017mvn} and \cite{Radice:2018pdn} where a comparison among all these numerical results is shown in their Table 3.  One can notice that the variability among the various estimates is rather large.
Concerning our work in particular, a few effects potentially relevant have not been incorporated 
such as a treatment of the neutrino transport, a fully consistent description of the thermal component of the EoS and also
we use a piece-wise polytropic approximation. 
However, our results are in a very good agreement with \cite{Shibata:2015prd}, where full beta equilibrium, thermal evolution and approximated neutrino cooling and absorption were considered.

Another important effect that one should consider is related to the dependence of $M_{\mathrm{ej}}$  and $M_{\mathrm{disk}}$ 
on the mass asymmetry $q$. This issue has been discussed in \citet{Rezzolla2010,Giacomazzo:2012zt} and more recently in \citet{Kiuchi:2019lls}. It is expected that $M_{\mathrm {disk}}$ is always larger for unequal-mass binaries because of the more efficient tidal interactions and angular momentum transfer during the merger. The same consideration seems not to apply in general for $M_{\mathrm{ej}}$
and further studies are needed in particular when considering models close to the threshold mass, because the state after the merger may rapidly change from producing a direct collapse to forming a short lived remnant \citep{Kiuchi:2019lls}.  

\subsection{Estimate of the Threshold mass}

A first estimate of the values of the threshold mass for the SFHo and 
SFHo-HD EoSs can be obtained by using the empirical formulae presented
in \citet{Bauswein:2013jpa,Bauswein:2015vxa,Bauswein:2017aur}.
The ratio $k=M_{\mathrm{threshold}}/\mtov$ scales linearly 
with the compactness of the maximum mass configuration $C_{max}$:
$k=2.43-3.38C_{max}$. In the case of SFHo and SFHo-HD, 
$C_{max}=0.3$ and $C_{max}=0.23$, respectively. 
Correspondingly, one obtains 
$M_{\mathrm{threshold}}=2.95\ms$  and $M_{\mathrm{threshold}}=2.61\ms$.
We can compare these values with the results that we obtain from our numerical simulations. Let us discuss first the case of SFHo-HD: 
from Table \ref{tab:MejMdisk} one can notice that for $\mtot >2.48\ms$ the remnant collapses in less than $1$ms 
and $M_{\mathrm{ej}}$ drops below m$\ms$. Moreover, there is almost no GW energy in the post merger phase. Thus from numerical simulations we infer for SFHo-HD that $M_{\mathrm{threshold}} \sim 2.5\ms$ (a few percent smaller than the estimate obtained with the empirical formula).  
Concerning SFHo, the largest value of $\mtot$ that we have simulated is $2.84 M_\odot$ and it leads to a a prompt collapse to a BH and represents indeed a good estimate of the 
threshold mass for the SFHo EOS (again a few percent smaller than the estimate obtained from the empirical relation).

\begin{figure}
	\begin{centering}
    	\epsfig{file=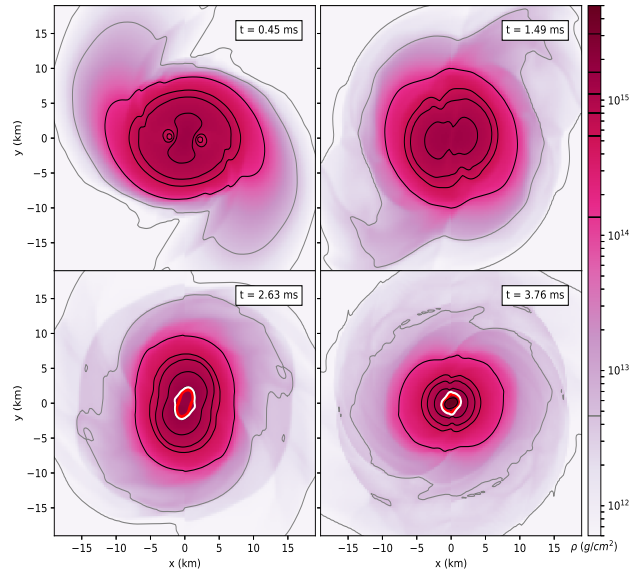,height=8.5cm,width=8.5cm}
        \caption{Projection in the XY plane of the density of the neutron star in the post-merger phase at the time of the peaks in the maximum baryon density. The model considered in this figure is the SFHo-HD with mass combination of $1.18 M_{\odot}$ - $1.18 M_{\odot}$. The thin black and gray contour lines correspond to the densities of the polytropic approximation used in this work. The thick red and white lines in the high-density region mark the region of the star in which quark nucleation can start (same region of the stripe displayed in Fig.8).}
        \label{fig:unstREGION} 
    \end{centering}
\end{figure}

\begin{figure}
	\begin{centering}
    	\epsfig{file=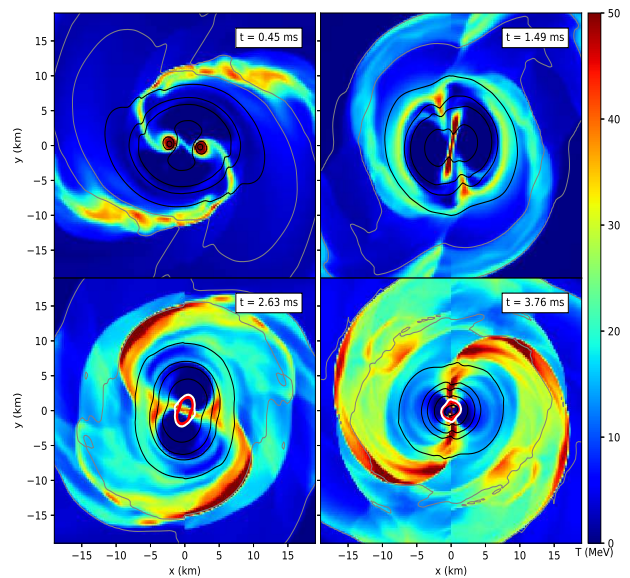,height=8.5cm,width=8.5cm}
        \caption{Projection in the XY plane of the temperature of the neutron star in the post-merger phase at the time of the peaks in the maximum baryon density. The model considered in this figure is the SFHo-HD with mass combination of $1.18 M_{\odot}$ - $1.18 M_{\odot}$. The thin black and gray contour lines correspond to the densities of the polytropic approximation used in this work. Thick red and white lines as in Fig.14.}
        \label{fig:tempREGION} 
    \end{centering}
\end{figure}

\subsection{Trigger of the phase conversion}

For $\mtot<\mth$, the two-families scenario predicts that
the phase conversion of hadronic matter 
to quark matter necessarily occurs. 
Simulating the process of conversion itself after the merger
is numerically very challenging (one needs to couple the code used for instance in \citet{Herzog:2011sn,Pagliara:2013tza} to \codename{Cactus}).
We can however estimate when the conversion should start.
As explained previously, the conversion is triggered only
when a significant amount of strangeness is produced during the evolution of the remnant, i.e. $Y_S \gtrsim 0.2$. 
In Fig.\ref{fig:BounceDP}, in the left panel,
we display the stripe in the mass density-temperature diagram for which the value of $Y_S$ is large enough
for the conversion to start. In the right panel
it is also shown the temporal evolution of the maximum baryon density for all the runs of HS-HS merger. Let us discuss first the case of the run SFHo-HD 118vs118.
One can see that  after the first two oscillations of the remnant, the condition for quark matter nucleation is fulfilled; see also Figs. \ref{fig:unstREGION} and \ref{fig:tempREGION} which allow to visualize the regions inside the remnant where the density and the temperature reach the condition
for the beginning of the conversion process.
The subsequent temporal evolution of the remnant would be a dramatic change of the structure of the star: within a few ms a big part of the star is converted into quark matter, the radius of the star increases by a few km and the heat released by the conversion heats up the star. Also, a significant amount of differential rotation can develop due to the very fast change of the total moment of inertia of the star \citep{Pili:2016hqo}. Finally, the stiffening of the EoS would stabilize the remnant with respect to the collapse.
We have a similar evolution for the SFHo-HD 12 model with the only difference that the conditions for the phase transition are met during the second bounce instead of the third one. 

A comment concerning the results for the mass ejected in these cases is in order. The oscillations of the maximum density are associated with the shock waves which in turn are responsible for most of the mass dynamically ejected.
Thus, even if we are not implementing in our simulations 
the conversion to quark matter (which could modify the temporal behaviour of the maximum density) we can assume that the results of the mass ejected by HS-HS mergers before the trigger of the phase transition (before the third peak in the SFHo-HD 118vs118 model) are fairly reliable.

Let us discuss now the cases with larger masses. 
For the model SFHo-HD 124vs124, which leads to a prompt collapse, the nucleation condition 
is met only when the collapse has already started and the formation of the quark phase cannot alt it. The intermediate case of SFHo-HD 122vs122 instead shows a first oscillation
during which the maximum density reaches the threshold for nucleation but the second oscillation leads already to the collapse of the star. In this case it is possible 
that a significant part of the star is converted to quark matter 
but it is quite likely that the consequent stiffening of the EoS is not strong enough to prevent the collapse.

\section{Rates of the mergers from population synthesis analysis}

\begin{table*}
\begin{center}
\begin{tabular}{|l||c|c|c||c|c|c||c|c|c|}
        \hline
 		&   \multicolumn{3}{c||}{all mergers} & \multicolumn{6}{c|}{GW170817-like}\\\hline
        & \multicolumn{3}{c||}{} & \multicolumn{3}{c||}{$0.7<q<1.0$} & \multicolumn{3}{c|}{$0.7<q<0.85$}\\\hline
	Model & HS-HS & HS-QS & QS-QS & HS-HS & HS-QS & QS-QS & HS-HS & HS-QS & QS-QS\\\hline
            \hline
            % Blanchard's SFH
%     $M_{\rm max}^{\rm H}=1.5\, M_\odot$ & 5.8 & 1.2 & 0.1 & 3.4 & 0.2 & 0.01 \\
%     $M_{\rm max}^{\rm H}=1.6\, M_\odot$ & 5.7 & 1.4 & 0.03 & 3.3 & 0.1 & 0.0 \\
%     one-family & 5.3 & -- & -- & 2.5 & -- & -- \\
			% Troja's SFH
    $M_{\rm max}^{\rm H}=1.5\, M_\odot$ & 9.1 & 3.1 & 0.2 & 6.4 & 0.4 & 0.01 & 
     0.03 & 0.2 & --\\\hline
    $M_{\rm max}^{\rm H}=1.6\, M_\odot$ & 9.2 & 3.2 & 0.02 & 6.5 & 0.3 & -- & 
      0.1 & 0.2 & --\\\hline
    one-family & 12.8 & -- & -- & 6.6 & -- & -- &
     0.3 & -- & --\\\hline
\end{tabular}
\end{center}
\caption{Merger rates from population synthesis [$\times10^{-3}{\rm yr}^{-1}$]. For the two families
scenario we have adopted two possible values of $M^{\mathrm{H}}_{\mathrm{\max}}$ and we report the results for all the merger combinations: HS-HS, HS-QS and QS-QS mergers. Results for the one-family scenario are also reported for comparison. Notice that for values of $q\lesssim 0.85$, a HS-QS merger is more probable than a HS-HS merger. }
\label{tab:popSyn} 
\end{table*}

The crucial information that we now want to provide is an estimate of the rate of events that we expect 
for the different types of mergers. For this purpose, we studied the evolution towards mergers of double compact objects in the two-families scenario. Specifically, we used the {\tt Startrack} population synthesis code \citep{Belczynski0206,Belczynski0801} with further updates described in \citet{Wiktorowicz1709} and references therein. As GW170817 is our main point of interest, we concentrated on systems having parameters similar to the observed ones and adapted the simulation parameters to reproduce an environment similar to the host galaxy (NGC4993).

Recently, \citet{Belczynski1807} performed a study of merger rates of two neutron stars 
in NGC4993 environment. Here, we include also QSs as possible components of the progenitor binary. Specifically, following \citet{Belczynski1807}, we assumed the metallicity of the environment in which GW170817 was formed to be $Z=0.01$, which corresponds to about $50\%$ of the solar metallicity. The results were scaled to the observable volume of aLIGO detector for events similar to GW170817, which may be represented approximately by a sphere with a radius of $D\approx100$ Mpc. On the base of Illustris simulation \citep{Vogelsberger1410}, this volume contains about $1.1\times10^{14} M_\odot$ of stellar mass in elliptical galaxies. The star-formation history was chosen to be uniform between $3$--$7$ Gyr ago \citep{Troja1711}, which is different to \citet{Belczynski1807}, who used a burst-like star-formation history occurring $1$, $5$, or $10$ Gyr ago.

As far as the binary evolution is concerned, all mass transfer was assumed to be conservative, i.e. all mass lost by the donor is transferred to the accretor. Such an approach increases the rates for double compact object mergers \citep{Chruslinska1803}. On the other hand, we kept the Maxwellian distribution of natal kicks to $\sigma=265$ km s$^{-1}$, which conforms to the observational estimates from Galactic's pulsars' proper motions \citep{Hobbs0507}, although lower natal kicks would increase the rates \citep{Chruslinska1803}.

Similar to \citet{Wiktorowicz:2017swq}, we adopted the two-families scenario by assuming that a HS converts into a QS when its gravitational mass reaches a maximal value, $M_{\rm max,ns}=M_{\rm max}^{\rm H}$.
During the transition, the gravitational mass of the compact object changes instantly (compared to the typical timescales of population synthesis analysis) and we update the orbital parameters accordingly.

Calculated merger rates are presented in Tab.~\ref{tab:popSyn}. Two models adopt the two-families scenario with deconfinement taking place at HS's mass of $1.5\ms$ and $1.6\ms$ (models $M_{\rm max}^{\rm H}=1.5\, M_\odot$ and $M_{\rm max}^{\rm H}=1.6\, M_\odot$, respectively). In the third model (one-family), which we provide for reference, all compact objects are HSs and the deconfinement process never occurs. We note that in all models it was assumed that all compact objects with mass above $M>2.5\ms$, formed prior to the merger, collapse to a BH and are not included in our study.

In our analysis we compare merger rates for all systems and merger rates corresponding to events similar to the GW170817 event.
The latter were chosen as events with chirp mass in the range $M_{\rm chirp}=(1.188\pm0.1) \ms$ (the $\pm0.1\ms$ was chosen arbitrarily, but the specific adopted value of the range has a negligible effect on the conclusions) and a mass ratio of $q>0.7$ ($q=m_2/m_1>0.7$, where $m_2\leq m_1$). 

Firstly, when considering all mergers, it is noticeable that the rate of HS-QS is not strongly suppressed in respect to the rate of HS-HS; they constitute roughly one fourth of the total rate.  

Secondly, concerning the GW170817-like events, different types of mergers are allowed depending on the value of the mass ratio $q$. In Tab.~\ref{tab:popSyn} we compare the merger rates for the entire range of mass ratios ($q>0.7$) and, separately, for lower mass ratios ($0.7<q<0.85$). The latter was motivated by an analysis of the GW170817 event \citep{Abbott:2018exr} indicating that a mass ratio  $q\sim0.85$ is highly compatible with the data. Results for both ranges of mass ratios are significantly different. The entire range ($q>0.7$) is dominated by high mass ratio systems which are mostly HS-HS, because for $q\approx1$ and $M_{\rm chirp}\approx1.188$ the components' masses can barely enter the QS mass range $M_{\rm min}^{\rm Q}>1.37\ms$, or $1.46\ms$ (for $M_{\rm max}^{\rm H}=1.5\, M_\odot$ and $M_{\rm max}^{\rm H}=1.6\, M_\odot$, respectively). On the other hand, for low mass ratios one of the stars usually becomes more massive than $M_{\rm min}^{\rm Q}$, thus, according to the two-families scenarios, it converts into a QS. As a result, although the mergers within the entire mass ratio range ($q>0.7$) show a preference for HS-HS mergers, the ones with a lower mass ratio ($0.7<q<0.85$) are typically HS-QS mergers. However, the former give much higher merger rates ($\sim6.8$ kyr$^{-1}$) than the latter ($\sim0.2$--$0.3$ kyr$^{-1}$).

It is also important to note that the rates of QS-QS mergers are in general very much suppressed, reaching at most a few percent of the total expected events involving at least one QS in the merger, a result in agreement with the findings of \citet{Wiktorowicz:2017swq}. In particular, the probability that GW170817 was due to a QS-QS merger is totally negligible.

Finally, we note that all the merger rates obtained from population studies are significantly below the one estimated from the GW170817 event 
(by more than two orders of magnitude), as previously shown by \citet{Belczynski1807}. It may be a result of, e.g. low observational statistics (just one event), poorly understood binary evolution phases (e.g. common envelope survival of low mass ratio binaries), or the existence of unknown evolutionary scenario.

In the following, we discuss the interpretations of GW170817 in terms of the different types of merger and their probability.

\subsection{HS-HS binary}

Our results show that, providing the entire possible range of mass ratio values ($q>0.7$) is taken into account, in the two-families scenario GW170817 is most probably a merger of two HSs. Indeed, while the observationally estimated chirp mass ($M_{\rm chirp}\approx1.188 M_\odot$) and mass ratio ($q>0.7$) limit the mass of the primary to $(1.37$--$1.64) M_\odot$ and of the secondary to $(1.17$--$1.37) M_\odot$, a result of population synthesis analysis is that mass ratios $q\sim 1$ are favoured and in equal mass binaries having $M\sim 1.37 M_\odot$ QSs are very unlikely to occur.

In a typical evolution, which produces a double HS merger in two-families scenario, the masses of the binary components on ZAMS are $9.7\ms$ and $8.6\ms$ and the separation is moderate ($a\approx340 R_\odot$). The primary, i.e. the heavier star on ZAMS, evolves faster and after $27$ Myr expands as a Hertzsprung Gap star, fills its Roche lobe and commences a mass transfer onto its companion. The mass transfer lasts $\sim20,000$ yr. Afterwards, the primary is ripped off its hydrogen envelope and left as a $2.1\ms$ Helium star. The companion accretes all of the transferred mass (as assumed in our computations) and reaches a mass of $16\ms$. After an additional $4$ Myr, the primary explodes as a SN and forms a $1.26\ms$ compact object. At that moment the separation is still high ($a\approx1900 R_\odot$). The secondary evolves and after an additional $4$ Myr expands and fills its Roche lobe as an AGB star. This time the mass ratio is too extreme ($q\approx0.24$) and the mass transfer is unstable leading to a common envelope. As a consequence, the separation shrinks to $a\approx 29 R_\odot$ and the secondary becomes a Helium star. Shortly after, it forms a compact object in a SN explosion. The natal kick slightly enlarges the orbit to $a\approx95 R_\odot$, but the eccentricity, $e\approx0.98$ is large enough to allow for a merger within $\sim5$ Gyr.

\subsection{HS-QS binary}

In the case of low mass ratios ($0.7<q<0.85$), the most probable
scenario involves the merger of a HS and a QS, because the most
massive component will typically be massive enough to become a
QS. This result is true in general and it also holds for GW170817-like
events. Notice though that for GW170817-like events the total merger
rate for low mass ratios ($\sim0.2$--$0.3$ kyr$^{-1}$) is more than an
order of magnitude smaller than for the wider range (i.e. $q>0.7$;
$\sim6.8$ kyr$^{-1}$).

In this situation, the progenitor binary on ZAMS typically consists of
$15.9\ms$ and $15.6\ms$ stars on an eccentric orbit ($e\approx0.8$)
with a separation of approximately $a\approx410 R_\odot$. The primary
fills its Roche lobe first and commences a mass transfer onto the
counterpart. The outcome is a mass reversal. The primary becomes a
$\sim4.4\ms$ Helium star, whereas the secondary is now a $26.5\ms$
main-sequence star. About $1$ Myr later, the primary forms a
$\sim1.2\ms$ NS in a SN explosion. The secondary quickly evolves,
expands, and fills its Roche lobe as a Hertzsprung Gap star. The donor
is much heavier than the accretor, so the mass flow becomes unstable
and results in a common envelope. The binary survives the phase but
separation is significantly smaller ($a\approx2 R_\odot$). The
secondary loses most of its mass and becomes a $\sim8\ms$ Helium star
which after a short evolution ($\sim1$ Myr) forms a compact object in
a SN and becomes a $\sim1.5\ms$ QS because it was too heavy to form a
HS. The separation grows to $\sim4.7 R_\odot$ due to natal kick, but
significant eccentricity allows for a merger within the next $\sim4.6$
Gyr.

\subsection{QS-QS binary}

As already discussed above, a GW170817-like merger is very unlikely to be of the QS-QS type. More precisely, GW170817 cannot be a QS-QS, as is immediately evident from Fig.~\ref{merger-cases}. GW170817-like mergers can marginally be QS-QS for $q\sim 1$ where they fit in the "all-comb" region of Fig.~\ref{merger-cases}.

\section{GW170817 and its interpretation as a HS-QS merger \label{gw17}} 

In the case of GW170817 the total mass is
$\mtot^{170817}=2.74^{+0.04}_{-0.01}\ms$ and, as explained before,
within the two-families scenario this event can be interpreted only as
due to the merger of a HS with a QS. Indeed $\mth<\mtot^{170817}$ and
therefore a HS-HS binary would promptly collapse to a BH. However, a
direct collapse is excluded by the observation of a sGRB $\sim 2$ sec
after the merger.  Also, as discussed previously it cannot be
interpreted as a merger of two QSs.

A first question concerns the probability of detecting such a HS-QS
merger event.  From the binary evolution point of view, the population
synthesis analysis has shown that the delay time, which is $\sim4.6$
Gyr, is consistent with the lack of recent star formation in the host
galaxy.

Concerning the phenomenology of GW170817, without detailed numerical
simulations of the merger of a HS with a QS it is difficult to firmly
establish whether in our scenario we can explain all the features of
the observed signals. We can however verify if our model satisfies the
upper and lower limits on $\tilde{\Lambda}$.  The lower limit on
$\tilde{\Lambda}$ in particular, empirically found in
\citet{Radice:2017lry}, stems from the request of describing the KN
signal AT2017gfo (this constraint has been recently criticized in
\citet{Kiuchi:2019lls}).

\begin{figure}%[!ht]
	\begin{centering}
		\epsfig{file=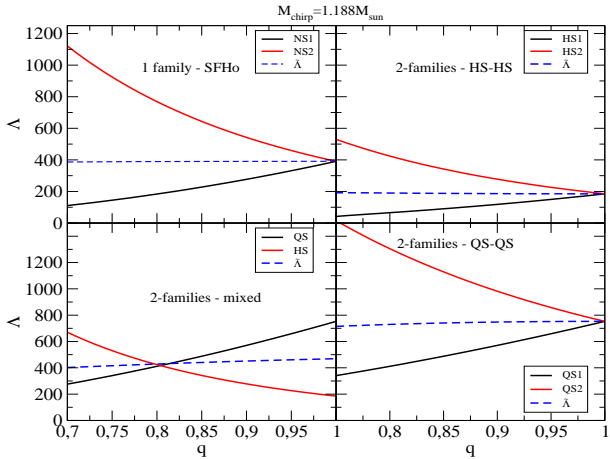,height=8cm,width=6cm,angle=-90}
		\caption{Tidal deformabilities for the components of a binary system (one family and two-families scenarios) for a value of the chirp mass equal to the one of GW170817.\label{lambdas} For the one family scenario we use the SFHo EoS, for the two families the SFHo-HD and the QS EoS and we consider the possibility of HS-HS, HS-QS or a QS-QS systems. }
	\end{centering}
\end{figure}

In Fig.\ref{lambdas}, we show the tidal deformabilities of the two components of the binary and $\tilde{\Lambda}$, for the chirp mass of GW170817, for the one-family and the two-families scenario (for the cases of HS-HS, HS-QS and QS-QS mergers \footnote{The QS-QS case is shown just for comparison but it 
cannot be realised because it violates the limit on $M^Q_{\mathrm{min}}$.}).
 Both the upper limit on $\tilde{\Lambda}$, as obtained by the GW's signal and the lower limit, as inferred from the KN analysis \citep{Radice:2017lry}, can be fulfilled within the two-families scenario in the case of a HS-QS system (see also \citet{Burgio:2018yix}).

We observe that the case of HS-HS is not compatible with the
constraints on $\tilde{\Lambda}$ because it provides a too small value
for it \citep{Radice:2017lry,Kiuchi:2019lls}.  Thus also from the
constraints on $\tilde{\Lambda}$, we can rule out the possibility that
GW170817 was due to a HS-HS merger.  We need to discuss now how the
electromagnetic counterparts of GW170817 would be produced in our
scheme.

Concerning the sGRB, we can assume that it has been produced by the
disk accreting onto the black hole in agreement with the
classification discussed in Sec.4. Indeed, for GRB170817A there is no
clear indication of the existence of an extended emission after the
prompt signal.  Concerning the KN, a crucial issue is connected with
the fate of the quark matter ejected from the QS during the merger.
There are a few studies indicating that quark matter evaporates into
nucleons and that this process can be so efficient that only
strangelets with baryon number much larger than $10^{40}$ can survive
\citep{Alcock:1985vc,Madsen:1986jg,drago19}.  Most of the matter
ejected from the QS will therefore rapidly evaporate into nucleons and
can contribute to the KN signal.  If this idea is correct, the KN
produced after a HS-QS merger would not show significant differences,
concerning the type of matter ejected, with respect to a KN produced
after a HS-HS merger.  Actually, the process of matter ejection can be
very efficient and it opens the possibility to explain AT2017 gfo as a
HS-QS merger: in the case of AT2017gfo the radii of the two compact
objects are both rather small, the system is asymmetric and the
threshold mass is large as discussed in Sec. 3. These are exactly the
requests posed in \citet{Kiuchi:2019lls} and, as discussed in that
paper, there are no examples of EoSs based on a single family scenario
able to satisfy all those requests.

Finally, we note that while the scenario of a delayed collapse (a few
tens of ms after the merger) is adopted by most of the literature on
GW170817, there are some studies indicating that the merger remnant
could have been active for a much longer time scale. In
\citet{vanPutten:2018abw}, a post-merger GW signal has been reported
with an associated GW energy lower than the sensitivity estimates of
the LIGO-VIRGO collaboration.  This signal would have lasted a few
seconds, with an initial frequency of about $700$ Hz.  Another
possible signal of a long lived remnant comes from the analysis of
\citet{Piro:2018bpl} in which it has been found a low-significance
temporal feature in the X-ray spectrum $\sim 150$ days after the
merger which is consistent with a sudden reactivation of the central
compact star.  If these analyses are correct, they can have very
important implications for the EoS of dense matter that we will
address in a forthcoming study.

\section{Discussion and Conclusions} \label{sec:conclusions}

We have explored the phenomenological consequences of the existence of
two families of compact stars on the observations of mergers. A
specific feature of the two-families scenario is the possibility that
at fixed values of the chirp mass and mass asymmetry, the merging
binary can be composed by two HSs, by two QSs or by a HS and a
QS. Each of these possibilities has its own specific signatures.

Let us first consider systems similar to the one associated with
GW170817.  Our population synthesis analysis basically excludes the
possibility that such binary systems are made of two QSs. They can be
formed either by two HSs or by a HS and a QS.

If we look to the whole range of values for the mass ratio $q$ of the
two stars the rate of HS-HS merger events is a factor $(16-20)$ larger
than the rate of HS-QS merger while for small mass ratio ($0.7 \leq q
\leq 0.85$) HS-QS systems are about 2-6 times more probable than HS-HS
systems.  While in the HS-HS case we would predict a prompt collapse
(no sGRB and a very faint KN), in the HS-QS case a hypermassive
remnant is more likely to form (with associated electromagnetic
signals similar to GRB170817A and AT2017gfo).

The smoking gun of our scenario would be just a single detection of a
source of GWs with a total mass smaller than the one of GW170817 but
lacking a significant electromagnetic counterpart (indeed it would be
interpreted as a HS-HS merger).  This possible signature is valid for
values of the total mass in the range $(2.48-2.74)\ms$ i.e. between
the threshold mass for the merger of two HSs and the mass of the
system associated with GW170817.

Another promising way to test the two-families scenario is related to
the observation of binaries with low values of $\mtot$. Let us
concentrate on $\mtot=2.4\ms$ with two stars of $1.2\ms$. A first
difference between the two-families and the one-family scenario
concerns the value of $\tilde{\Lambda}$ which is significantly smaller
(almost a factor of 2) for a merger of HS-HS with respect to the
merger of two NSs (see the values for SFHo-HD and SFHo in Table
1). This is due to the appearance of delta resonances in hadronic
matter at densities of about twice the saturation density.  Accurate
future measurements of the GW's signal associated with the inspiral
phase should put interesting upper limits on $\tilde{\Lambda}$ and
test therefore the possibility of merger of very compact stellar
objects, i.e. HSs in our scheme.

Additionally, if observations of the postmerger signal will be
feasible by future experiments (most probably the third generation
detectors), one can test the two families scenario through the
measurement of the frequency of the $f_2$ mode which for a HS-HS
merger is of the order of 1kHz larger than the case of a NS-NS merger
during the first milliseconds of the life of the remnant.  The
subsequent formation of quark matter, which in our scenario
corresponds to a stiffening of the EoS, should then decrease the value
of $f_2$ (see \citet{Bauswein:2015vxa} for a preliminary analysis of
this signature). It is interesting to note that in recent studies
discussing a phase transition to quark matter in the postmerger, the
appearance of quark matter reduces the lifetime of the remnant and
leads to a shift of $f_2$ to frequencies larger than the ones of the
"progenitor" made only of nucleonic matter
\citep{Most:2018eaw,Bauswein:2018bma}.

Let us discuss what are the possible signatures associated with the
KN. Its signal depends strongly on the amount of mass dynamically
ejected by the merger and on the amount of mass in the disk. One can
notice from Table \ref{tab:MejMdisk} that for a HS-HS merger both
these masses are of the order of $0.01\ms$. Instead in the case of a
NS-NS merger (in the one-family scenario) the mass dynamically ejected
is significantly smaller than the mass of the disk.  Qualitatively, we
would then expect a significant difference between the resulting KN
signals: within the two-families scenario (in the case of a HS-HS
merger) the intermediate opacity purple component is strongly
suppressed with respect to the components associated to the
dynamically ejected matter \citep{Perego:2017wtu}.

Finally, it is interesting to notice that the suggestion of the
two-families scenario originated from the requirement of explaining
the observations of stars having large masses and the possible
existence of stars with small radii. The same problem presents itself
again when trying to model the KN AT2017gfo: as suggested in
\citet{Siegel:2019mlp}, the presence of a strong shock-heated
component in the ejecta would favor radii smaller than 11-12km. Such
small radii can be easily accommodated in the two-families scenario but
are basically ruled out within the one-family scenario.

%%%%%%%%%%%%%%%%%%%%%%%%%%%%%%%%%%%%%%%%%%%%%%%%%%%%%%%%%%%%%%%%%%%%
\vskip 1cm 
We would like to thank thousands of volunteers who by their
participation in the Universe@Home
project\footnote{\url{http://universeathome.pl}} performed calculation
necessary for the population synthesis studies.  We benefited from the
availability of public software that enabled us to conduct the 3D
general relativistic simulations, namely ``LORENE'' and the ``Einstein
Toolkit''. We do express our gratitude to the many people that
contributed to their realization. This work would have not been
possible without the CINECA-INFN agreement that provides access to
resources on GALILEO and MARCONI at CINECA.  We acknowledge PRACE for
awarding us access to MARCONI at CINECA, Italy under grant
Pra14\_3593. G.W. was partly supported by the President’s
International Fellowship Initiative (PIFI) of the Chinese Academy of
Sciences under grant no.2018PM0017 and by the Strategic Priority
Research Program of the Chinese Academy of Science Multi-waveband
Gravitational Wave Universe (Grant No. XDB23040000). Support from INFN
``Iniziativa Specifica NEUMATT'' is kindly acknowledged.

\bibliographystyle{apj}
\bibliography{references}

\end{document}